\PassOptionsToPackage{dvipsnames}{xcolor}

\documentclass[acmsmall,screen]{acmart}
\setcopyright{cc}
\setcctype{by}
\acmDOI{10.1145/3808263}
\acmYear{2026}
\acmJournal{PACMPL}
\acmVolume{10}
\acmNumber{PLDI}
\acmArticle{185}
\acmMonth{6}
\acmSubmissionID{pldi26main-p56-p}
\received{2025-11-14}
\received[accepted]{2026-04-03}

\usepackage{booktabs}   
\usepackage[normalem]{ulem}
\usepackage{verbatim}
\usepackage[T1]{fontenc}
\usepackage{epsfig,endnotes,footnote}
\usepackage{array}
\usepackage{multirow}
\usepackage{amsmath}
\usepackage{dsfont}
\usepackage{enumitem}
\usepackage{xspace}
\usepackage{relsize}
\usepackage{hyphenat}
\usepackage{algorithm}
\usepackage{algorithmicx}
\usepackage{algpseudocode}
\usepackage{rotating}
\usepackage{graphicx}
\usepackage{subfig}
\DeclareGraphicsExtensions{.pdf,.jpg,.png}

\usepackage{url}
\usepackage{wrapfig}
\usepackage{pifont}
\usepackage{cleveref}
\usepackage{tabularx}
\usepackage{mathpartir}
\usepackage{colortbl}
\usepackage{listings}
\usepackage{esvect}

\hypersetup{
  unicode=false,          
  pdftoolbar=true,        
  pdfmenubar=true,        
  pdffitwindow=true,      
  pdftitle={Let It Flow: A Formally Verified Compilation Framework for Asynchronous Dataflow},
  pdfauthor={Zhengyao Lin, Yi Cai, and Milijana Surbatovich},
  pdfsubject={Formally verified compilation for asynchronous dataflow},
  pdfnewwindow=true,      
  pdfkeywords={compiler verification, capability type system, dataflow architectures, coarse-grained reconfigurable arrays, high-level synthesis},
  colorlinks=true,        
  linkcolor=blue,         
  citecolor=blue,         
  filecolor=magenta,      
  urlcolor=cyan           
}

\Crefname{figure}{Fig.}{Figs.}
\Crefformat{section}{\S#2#1#3}
\Crefformat{subsection}{\S#2#1#3}
\Crefformat{subsubsection}{\S#2#1#3}
\Crefname{lstnumber}{Line}{Lines}
\let\autoref\Cref

\newcommand{\system}{Wavelet\xspace}

\newcommand{\lzero}{\ensuremath{\mathbb L^*_\mathit{let}}\xspace}
\newcommand{\lzeroghost}{\ensuremath{\mathbb L_\mathit{let}}\xspace}
\newcommand{\lone}{\ensuremath{\mathbb L_\mathit{flow}}\xspace}

\newcommand{\notes}[2]{} 

\newcommand{\extra}[1]{}

\newcommand{\m}[1]{\mathsf{#1}}
\newcommand{\mt}[1]{\mathit{#1}}

\makeatletter
\newcommand\paragraphb[1]{\def\@toclevel{4}%
  \@startsection{paragraph}{4}{\z@}%
  {-.3\baselineskip \@plus -2\p@ \@minus -.2\p@}%
  {-3.5\p@}%
  {\ACM@NRadjust{\@parfont}\bfseries}{#1.}}
\makeatother
\newcommand{\bnfdef}{::=}
\newcommand{\bnfalt}{\,|\,}

\newcommand{\checkcolor}{\textcolor[RGB]{57,124,43}}
\newcommand{\shrdtp}{\m{shrd}}
\newcommand{\uniqtp}{\m{uniq}}
\newcommand{\capsep}{\cdot}
\newcommand{\capsym}{C}
\newcommand{\fence}{\texttt{---}}
\newcommand{\judg}[5]{#1 ; #2 \;\vdash_{#3}\; #4 : #5}
\newcommand{\env}{\Gamma}
\newcommand{\capa}{\Delta}
\newcommand{\cond}{\Phi}

\newcommand{\vx}{\vv{x}}
\newcommand{\vy}{\vv{y}}
\newcommand{\vi}{\vv{i}}

\newcommand{\shrd}[1]{\mathsf{shrd@}\{#1\}}
\newcommand{\uniq}[1]{\mathsf{uniq@}\{#1\}}

\newcommand{\fnctx}{\Pi}

\newcommand{\aperm}{\setlength{\fboxsep}{1pt}\colorbox[RGB]{33,252,253}}
\newcommand{\bperm}{\setlength{\fboxsep}{1pt}\colorbox[RGB]{224,230,132}}
\newcommand{\ite}[3]{\ensuremath{\mathbf{if}~#1~\mathbf{then}~#2~\mathbf{else}~#3}}
\newcommand{\letin}[3]{\ensuremath{\mathbf{let}~#1 = #2;~#3}}
\newcommand{\ret}{\ensuremath{\mathbf{ret}}}
\newcommand{\tailcall}{\ensuremath{\mathbf{tail}}}

\newcommand{\para}{\ensuremath{\ ||\ }}

\newcommand{\apmerge}{\ensuremath{\mathsf{merge}}\xspace}
\newcommand{\apsteer}{\ensuremath{\mathsf{steer}}\xspace}
\newcommand{\apfork}{\ensuremath{\mathsf{fork}}\xspace}
\newcommand{\aporder}{\ensuremath{\mathsf{order}}\xspace}
\newcommand{\apsink}{\ensuremath{\mathsf{sink}}\xspace}
\newcommand{\apforward}{\ensuremath{\mathsf{forward}}\xspace}
\newcommand{\apconst}{\ensuremath{\mathsf{const}}\xspace}
\newcommand{\apop}{\ensuremath{\mathsf{op}}\xspace}

\newcommand{\configlzero}{C^\mathit{let}}
\newcommand{\configlone}{C^\mathit{flow}}

\newcommand{\yieldl}[1]{\mathsf{yield}(#1)}
\newcommand{\inputl}[1]{\mathsf{input}(#1)}
\newcommand{\outputl}[1]{\mathsf{output}(#1)}

\newcommand{\interp}[1]{\llbracket #1 \rrbracket}
\newcommand{\lts}[2]{\mathsf{LTS}^{#2}_{#1}}
\newcommand{\labelset}{\mathit{Label}}

\newcommand{\push}{\ensuremath{\mathsf{push}}}
\newcommand{\pop}{\ensuremath{\mathsf{pop}}}

\newcommand{\iosim}{\ensuremath{\lesssim}}

\newcommand{\cfc}{\ensuremath{\mathsf{cfc}}\xspace}
\newcommand{\comp}{\ensuremath{\mathsf{comp}}\xspace}

\newcommand{\linksem}{\ensuremath{\mathsf{link}_\mathit{sem}}}
\newcommand{\linksyn}{\ensuremath{\mathsf{link}_\mathit{syn}}}

\newcommand{\spec}{\ensuremath{\mathsf{spec}}\xspace}
\newcommand{\precond}{\ensuremath{\mathsf{pre}}\xspace}
\newcommand{\postcond}{\ensuremath{\mathsf{post}}\xspace}

\newcommand{\guard}{\ensuremath{\mathsf{guard}}\xspace}
\newcommand{\triv}{\ensuremath{\mathsf{triv}}\xspace}

\newcommand{\join}{\ensuremath{\mathsf{jnsplt}}\xspace}

\newcommand{\ext}{\ensuremath{\mathsf{ext}}}

\newcommand{\guardrel}{\ensuremath{\triangleright}\xspace}

\usepackage{eqparbox}
\newdimen{\algindent}
\algnewcommand\LeftComment[2]{%
\hspace{#1\algindent}$\triangleright$ \eqparbox{COMMENT}{#2} \hfill %
}

\newif\ifproofs
\proofstrue

\usepackage{listings}

\definecolor{L0Keyword}{RGB}{0, 0, 0}        
\definecolor{L0Type}{RGB}{0, 0, 0}            
\definecolor{L0Operator}{RGB}{0, 0, 0}        
\definecolor{L0Comment}{RGB}{0, 0, 0}         
\definecolor{L0String}{RGB}{0, 0, 0}          

\lstdefinelanguage{L0}{
  morekeywords={
    def,
    let,
    if,
    then,
    else,
    in,
    end,
    return
  },
  morekeywords=[2]{
    u8,
    u16,
    u32,
    u64,
    i8,
    i16,
    i32,
    i64,
    bool,
    unit
  },
  morekeywords=[3]{
    true,
    false,
    load,
    store,
    ld,
    st,
  },
  sensitive=true,
  morecomment=[l]{--},
  morecomment=[n]{/*}{*/},
  morestring=[b]{"},
  morestring=[b]{'},
  morestring=[d]{\%},
  literate={
    {->}{$\to$}{1}
    {=>}{$\Rightarrow$}{1}
    {|-}{$\vdash$}{1}
  }
}[keywords, comments, strings]

\newcommand{\Lzerocode}[1]{\lstinline[language=L0]|#1|}

\definecolor{GhostKeyword}{RGB}{0, 0, 0}        
\definecolor{GhostType}{RGB}{0, 0, 0}           
\definecolor{GhostPermission}{RGB}{0, 0, 0}     
\definecolor{GhostComment}{RGB}{34, 139, 34}  

\lstdefinelanguage{L0Ghost}{
  morekeywords={
    def,
    let,
    if,
    then,
    else,
    in,
    end,
    jnsplt,
    const,
    op,
    ret,
    return
  },
  morekeywords=[2]{
    int,
    bool,
    unit
  },
  morekeywords=[3]{
    true,
    false,
    load,
    store,
    ld,
    st,
  },
  sensitive=true,
  morecomment=[l]{--},
  morecomment=[n]{/*}{*/},
  morestring=[b]{"},
  morestring=[b]{'},
}[keywords, comments, strings]

\lstdefinestyle{l0Ghost}{
  language=L0Ghost,
  basicstyle=\ttfamily\small,
  keywordstyle=\bfseries\color{GhostKeyword},
  keywordstyle=[2]\bfseries\color{GhostType},
  keywordstyle=[3]\bfseries\color{GhostPermission},
  commentstyle=\itshape\color{GhostComment},
  stringstyle=\color{L0String},
  mathescape=true,
  breaklines=true,
  breakatwhitespace=true,
  tabsize=2,
  showstringspaces=false,
  xleftmargin=0pt,
  xrightmargin=0pt,
  numbers=none,
  frame=single,
  framexleftmargin=2pt,
  framextopmargin=2pt,
  framexbottommargin=2pt,
  framexrightmargin=2pt,
  backgroundcolor=\color{white},
  rulecolor=\color{black},
  captionpos=b,
  aboveskip=0.8\baselineskip,
  belowskip=0.6\baselineskip
}

\lstdefinelanguage{Lean4}{
  sensitive=true,
  alsoletter={\?,_},
  morekeywords={
    theorem,lemma,example,def,abbrev,axiom,inductive,structure,class,instance,
    namespace,section,end,open,variable,variables,universe,universes,
    mutual,where,deriving,local,private,protected,scoped,macro,syntax,notation,
    infix,infixl,infixr,prefix,postfix,alias,attribute,set_option,
    noncomputable,unsafe,partial,termination_by,decreasing_by
  },
  morekeywords=[2]{
    by,calc,have,show,from,let,in,fun,match,with,if,then,else,do,return,
    forall,forall',intro,intros,refine,apply,exact,assumption,cases,rcases,
    rfl,rewrite,rw,simp,dsimp,unfold,rename_i,decide,admit,sorry
  },
  morekeywords=[3]{Type,Prop,Sort,Nat,Int,Bool,Float,String,List,Option,IO},
  morecomment=[l]{--},
  morecomment=[s]{/-}{-/},
  morestring=[b]",
  morestring=[b]',
  literate=
   {->}{{$\to$}}2 {<-}{{$\leftarrow$}}2 {=>}{{$\Rightarrow$}}2
   {<->}{{$\leftrightarrow$}}3 {<=>}{{$\Leftrightarrow$}}3
   {/\\}{{$\wedge$}}2 {\\/}{{$\vee$}}2 {~}{{$\neg$}}1
   {forall}{{$\forall$}}7 {exists}{{$\exists$}}6
   {<=}{{$\le$}}2 {>=}{{$\ge$}}2 {!=}{{$\ne$}}2
   {λ}{{$\lambda$}}1 {→}{{$\to$}}1 {←}{{$\leftarrow$}}1 {↔}{{$\leftrightarrow$}}1
   {⟨}{{$\langle$}}1 {⟩}{{$\rangle$}}1 {⟨⟩}{{$\langle\rangle$}}2
   {×}{{$\times$}}1 {∘}{{$\circ$}}1 {⊢}{{$\vdash$}}1
   {∧}{{$\wedge$}}1 {∨}{{$\vee$}}1 {¬}{{$\neg$}}1
   {≤}{{$\le$}}1 {≥}{{$\ge$}}1 {≠}{{$\ne$}}1
   {⊆}{{$\subseteq$}}1 {⊂}{{$\subset$}}1 {∈}{{$\in$}}1 {∉}{{$\notin$}}1
   {⊥}{{$\bot$}}1 {⊤}{{$\top$}}1
   {α}{{$\alpha$}}1 {β}{{$\beta$}}1 {γ}{{$\gamma$}}1 {δ}{{$\delta$}}1
   {ε}{{$\varepsilon$}}1 {φ}{{$\varphi$}}1 {Γ}{{$\Gamma$}}1 {Δ}{{$\Delta$}}1
  ,
  keepspaces=true,
  showstringspaces=false,
  columns=fullflexible,
  upquote=true
}

\lstdefinestyle{lean4}{
  language=Lean4,
  basicstyle=\ttfamily\small,
  keywordstyle=\bfseries,
  keywordstyle=[2]\bfseries,
  keywordstyle=[3]\bfseries,
  commentstyle=\color{gray!65!black},
  stringstyle=\color{orange!70!black},
  numberstyle=\scriptsize\color{gray!70!black},
  numbers=left,
  numbersep=8pt,
  frame=single,
  framerule=0.4pt,
  rulecolor=\color{black!12},
  tabsize=2,
  aboveskip=0.8\baselineskip,
  belowskip=0.6\baselineskip
}

\begin{document}

\title{Let It Flow: A Formally Verified Compilation Framework for Asynchronous Dataflow}

\author{Zhengyao Lin}
\orcid{0000-0001-5475-5765}
\affiliation{
  \institution{Carnegie Mellon University}
  \city{Pittsburgh}
  \country{USA}
}
\email{zhengyal@cmu.edu}

\author{Yi Cai}
\orcid{0009-0001-9810-4048}
\affiliation{
  \institution{University of Maryland, College Park}
  \city{College Park}
  \country{USA}
}
\email{yicai@umd.edu}

\author{Milijana Surbatovich}
\orcid{0009-0004-6948-6683}
\affiliation{
  \institution{University of Maryland, College Park}
  \city{College Park}
  \country{USA}
}
\email{milijana@umd.edu}



\begin{abstract}

Dataflow architectures have gained renewed interest due to their balance between energy
efficiency and performance. In (spatial) dataflow architectures, a program is represented
as a set of entirely distributed and dynamically scheduled dataflow operators that
communicate through asynchronous channels, which greatly improves data locality and
parallelism. However, compiling to dataflow architectures remains an error-prone process,
due to the difficulty of maintaining \emph{determinacy} while enabling \emph{pipelining}.
Determinacy means that the result of a dataflow program is deterministic and independent
of the schedule of operator execution, and pipelining is an important optimization in
spatial dataflow that enables parallelism across loop iterations.

In this work, we present Wavelet, the first effort to formally verify a compiler for
asynchronous dataflow. We use a mix of techniques to achieve this goal. Our frontend
uses a novel capability type system with fences to synchronize conflicting memory
accesses and enable \emph{pipelining}. We then verify a Lean formalization of two
core compiler passes that translate elaborated programs from the type checker to
dataflow graphs, proving important properties of \emph{forward simulation} and
\emph{determinacy}. Notably, our formalization semantically propagates the soundness
guarantees of the frontend type system, ensuring modularity between simulation and
determinacy proofs. In our evaluation, we show that dataflow graphs compiled by
Wavelet have comparable quality to those produced by unverified dataflow compilers
from RipTide and LLVM CIRCT.

\end{abstract}

\maketitle
\hypersetup{pdfauthor={Zhengyao Lin, Yi Cai, and Milijana Surbatovich}}

\section{Introduction}
\label{sec:intro}
Challenges in decreasing transistor size, memory access bottlenecks, and balancing power and cooling all hinder
processor efficiency. Application domains that demand 
extreme power efficiency, e.g., from ultra-low-power edge computing~\cite{snafu} 
to high-performance machine learning~\cite{plasticine}, require hardware designers to look beyond traditional ``von Neumann'' architecture designs. One attractive alternative is the \emph{spatial dataflow architecture},
which has seen a recent resurgence in interest~\cite{riptide, pipestitch, marionette, plasticine, ripple, tyr, sam}, 
due to its outstanding balance of power efficiency and performance.

Unlike a traditional von Neumann architecture, where instructions are fetched 
and executed seemingly according to program order, in spatial dataflow architectures, 
instructions in a program are compiled to a
set of entirely distributed dataflow \emph{operators} connected with asynchronous channels.
At the hardware level, these dataflow operators are mapped to a physical array of ``processing elements'' (PEs), each serving as an operator (e.g., an arithmetic operation, 
a memory access operation, or a control-flow operation).
These PEs communicate through an on-chip, low-latency 
network, enabling direct dataflow between operators and significantly reducing data movement and false serialization points compared to traditional von Neumann architectures.

Historically, dataflow computing has seen challenges in maintaining programmability~\cite{ttda,wavescalar,yazdanpanah-dataflow},
only mapping small fragments of kernels, or being implemented in brittle, fixed-function circuits. 
Recent work on RipTide~\cite{riptide}
has enabled general-purpose programming while retaining efficiency, greatly increasing the viability of 
dataflow computing.
Unfortunately, while the architecture thus supports 
general-purpose and efficient computing, actually developing the language and compiler
infrastructure to realize this potential remains challenging.
Since a dataflow program executes as a distributed network of operators, extreme care must be taken 
to avoid bugs like racing parallel accesses to shared memory, or deadlocks due to a set of operators each waiting on the others to produce the 
data needed to execute. Indeed, because programming at the low level of dataflow graphs is difficult and unintuitive, 
in the standard existing workflow, programmers write applications in imperative, \emph{sequential} languages 
like C, relying on a compiler to translate their traditional sequential program to the distributed, parallel dataflow program. While such a workflow improves programmability, the 
dataflow compiler is left with the complex task of ensuring memory 
consistency and equivalence,
requiring layers of static analyses to ostensibly convert the original imperative program to an equivalent distributed one, particularly one that remains efficient. 

Thus, correct compiler design is an integral, foundational topic in the (re-)emerging area of dataflow architectures.
In particular, an ideal compiler for dataflow architectures should produce \emph{determinate} and \emph{pipelined} dataflow graphs:
determinacy means that the result of a dataflow graph is deterministic and independent
of the schedule of operator execution, precluding the possibility of data races. 
Meanwhile, pipelining is one of the most important optimizations that utilizes the parallelism in spatial dataflow, where computation of future loop iterations can start before the previous iteration has finished.
In essence, pipelining requires removal of unnecessary dependencies between operations, but doing so risks introducing data races and violating determinacy.

We argue that \emph{formal verification} is a crucial step in developing dataflow compilers that generate correct and efficient programs.
Rather than relying on more obviously correct 
but overly conservative compiler behavior,
we should design compilers that are mathematically guaranteed 
to generate dataflow programs that preserve the intended behavior of the sequential application written by 
the programmer, while not restricting desired parallelism unless needed for correctness. 

To that end, we present \system, the first formally verified compiler for asynchronous dataflow.
To approach the challenges of determinacy and pipelining, we use a combination of typing and compiler verification techniques.
Our core compiler passes that translate from an intermediate sequential language \lzeroghost to a dataflow calculus \lone are formally verified in the Lean 4 theorem prover~\cite{lean4}, guaranteeing forward simulation and deadlock-freedom.
To enable pipelining, our frontend language is equipped with a novel \emph{capability type system with fences} to ensure consistent usage of shared memory while not over-synchronizing memory operations.
The two layers of the compiler work together to achieve determinacy: the type system elaborates memory synchronization with \emph{affine permission tokens}, and our compiler proofs propagate the typing guarantees through simulation proofs, proving the determinacy of the final dataflow program.

Our compiler proofs are designed to be modular and extensible.
By interpreting \lzeroghost and \lone programs in a common semantic framework, we modularly verify the compilation of a single function and then extend the proof to whole programs using a separate proof for linking dataflow graphs.
By formulating the soundness condition of affine permission tokens as label restrictions, we carry the typing guarantees through multiple simulation passes \emph{without modifying them}, and then prove determinacy independently from forward simulations.

In summary, our main contributions are:
\begin{itemize}
    \item A capability-based type system with fences to ensure race-freedom and facilitate pipelining.
    \item Formally verified core parallelization passes of a dataflow compiler in Lean, guaranteeing forward simulation and determinacy.
    \item A novel technique of semantic affine permissions to propagate the frontend typing soundness guarantees throughout simulation and determinacy proofs, greatly improving modularity.
\end{itemize}

\begin{figure*}[tb]
    \centering
    \includegraphics[width=\columnwidth]{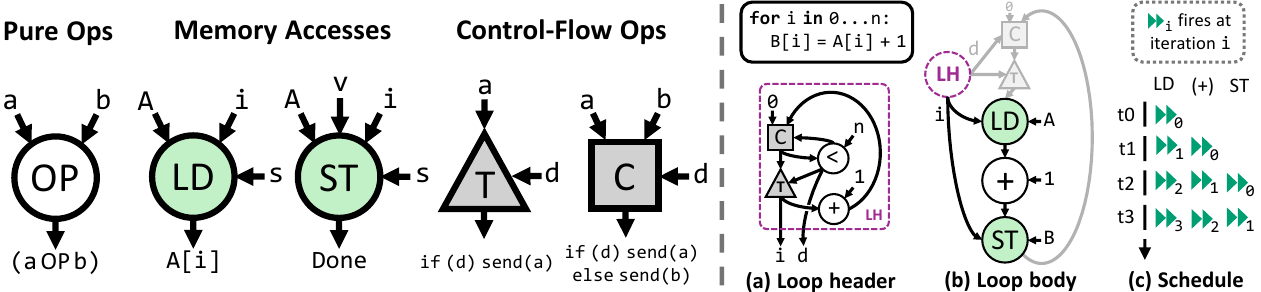}
    \caption{Overview of key dataflow operators and an example dataflow graph with its pipelined schedule.}
    \Description{Overview of key dataflow operators and an example dataflow graph with its pipelined schedule.}
    \label{fig:df-ops}
\end{figure*}

\section{Background \& Motivation}
\label{sec:background}
We explain the basics of compilation for dataflow architectures and the challenges of maintaining 
correct yet efficient execution, motivating the design of our verified compilation framework. 

\paragraphb{Dataflow Operators}
A dataflow compiler translates code written in a sequential language to 
a \emph{dataflow program}. These programs are represented as dataflow graphs, where the nodes of the graph are \emph{operators} and the edges are communication \emph{channels} by which data flows between operators. 
The set of available operators essentially forms the ISA of a particular dataflow architecture. 
We show representative operators in \autoref{fig:df-ops}, based on the RipTide ISA~\cite{riptide}. 
A pure \textsf{OP} (arithmetic/logic: $+$, $<$, etc.) consumes inputs from channels $a$ and $b$ and outputs the operation's result (e.g., $a + b$).
Memory operators like load ($\m{LD}$) and
store ($\m{ST}$) may modify main memory state. 
Both take a base address $A$, an index $i$, and an optional control signal $s$. When
all inputs arrive, $\m{LD}$ outputs the value at $A[i]$; $\m{ST}$
stores input $v$ at $A[i]$ and optionally outputs a \texttt{Done} control signal, which is critical for enforcing memory ordering.
Because operators fire once inputs are ready, (conflicting) memory operations might execute out-of-order.
To prevent this, compilers connect the \texttt{Done} signal of a preceding $\m{ST}$ to the signal input $s$ of a subsequent $\m{LD}$ or $\m{ST}$.
This explicit data dependency prevents the subsequent operation from firing prematurely, transforming implicit sequential memory ordering into explicit dataflow. 
Finally, control flow is implemented by routing values conditionally. 
A steer operator (\textsf{T}) acts as a branch filter: given decider $d$ and input $a$, it forwards $a$ if $d$ is true, and \emph{discards} it otherwise.
A carry operator (\textsf{C}) acts as a loop variable: it initially forwards its left input $a$ (the loop's initial value), then on subsequent iterations forwards whichever input decider $d$ selects.

\paragraphb{Dataflow Programs}
We now step through an example dataflow program (right side of \autoref{fig:df-ops}) which increments each element of an array $A$ and stores the result in $B$.
We show the loop header and body subgraphs in separate columns (a) and (b) for clarity.
The loop header in column (a) is responsible for generating a stream of values $0, \dots, n - 1$ for the loop variable $i$.
It starts by emitting the initial value $0$ from the carry operator, which is then passed to the comparison operator to check the loop condition $i < n$.
If this check passes, the steer operator then allows the value of $i$ to flow to both the add operator and the rest of the loop body.
The add operator increments $i$ and feeds the new value back to the $\m{carry}$ through a back-edge, and the same process repeats until the loop condition fails.
Meanwhile, the loop body in column (b) takes a stream of values for $i$ from the loop header, and performs the load of $A[i]$, the increment $A[i] + 1$,
and the store to $B[i]$ for each $i$.

Notice that the graph edges in column (a) form a cycle; $i$ 
has a loop-carried dependence, requiring its updated value to be passed back to the carry operator on each loop iteration. On the other hand, 
depending on whether $A$ and $B$ are disjoint, the loop body may have no such dependencies, indicating that iterations of the body 
can be parallelized, as we explain next.

\paragraphb{Pipelining and Determinacy}
Pipelining is a key optimization in dataflow architectures that enables more parallelism.
In \autoref{fig:df-ops}(b),
the grayed-out subgraph introduces a back-edge from the store to the load. 
If we do \emph{not} have this back-edge, the dataflow graph is \emph{pipeline-able}, 
i.e., the load in iteration $i + 1$ does not have to wait for the store in iteration $i$ to finish.
More specifically, consider the schedule of loop body operators in column (c).
The horizontal axis represents the three operators in the loop body ($\m{LD}$, $+$, and $\m{ST}$), and the vertical axis represents abstract time steps.
Assume that the loop header can stream one value of $i$ out per time step, with the load of 
iteration 0 occurring at time $t_0$.
At time $t_1$, the load of iteration 1 can run in parallel with the add of iteration 0;
by time $t_3$, the three operators in the loop body are simultaneously carrying out computation from three different iterations.
At the hardware level, this ideal situation is complicated by the actual timing of each operator, 
but at the compiler level, we want to maximize such abstract pipelining by not adding unnecessary data dependencies such as the gray part in column (b).
A pipelined dataflow graph carries a higher risk of losing \emph{determinacy}, however.
For example, the loop body is reading from $A$ and writing to $B$.
If $A$ and $B$ might alias, then the schedule in column (c) could cause data races and non-deterministic results 
(e.g., if the store of $B[0]$ and load of $A[2]$ at time $t_2$ are accessing the same location).
In this case, the gray back-edge becomes necessary to avoid data races and preserve the original memory dependency in the source sequential program.
Therefore, a key challenge in \system is to enable pipelining by eliminating unnecessary memory dependencies through a type system, while preserving correctness and determinacy with formalized and mechanized proofs.

\begin{figure}[tb]
    \centering
    \includegraphics[width=\textwidth]{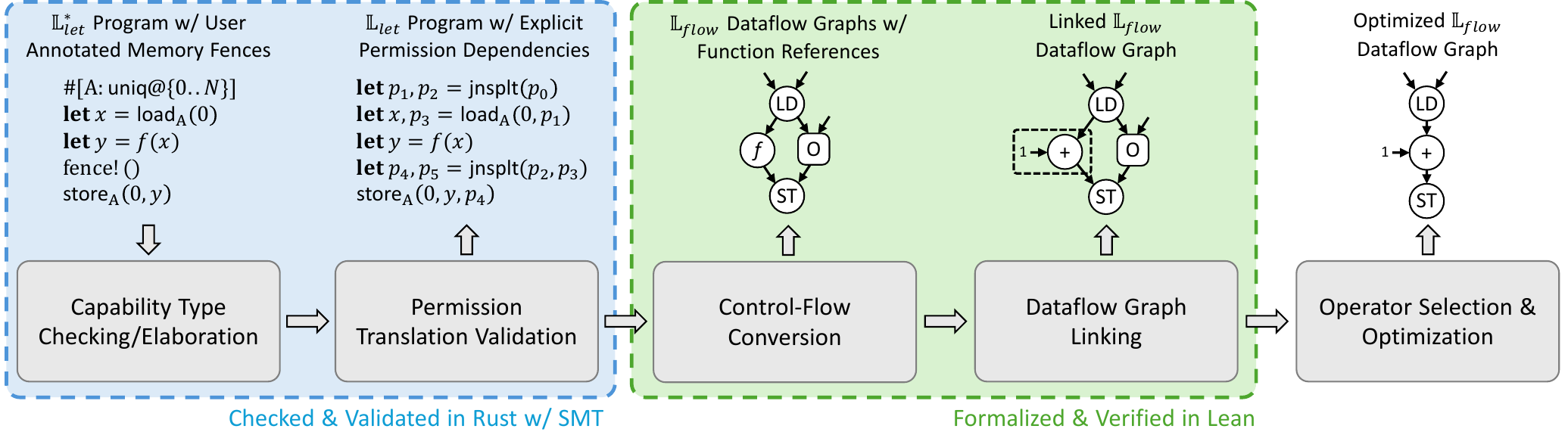}
    \caption{Compiler pipeline of \system. 
The compilation
transforms the frontend language \lzero through the elaborated IR \lzeroghost to
the dataflow calculus \lone.}
    \Description{Compiler pipeline.}
    \label{fig:pipeline}
\end{figure}

\section{System Overview}
\label{sec:overview}

We present an overview of \system's compilation pipeline in \autoref{fig:pipeline}.
At a high level, \system takes in a
sequential program on the left side, decorated with \emph{capability types}
(\texttt{\#[A: uniq@\{0..N\}]}) for accessing
arrays and user-inserted \emph{fences} (\texttt{fence!()}) indicating synchronization points,
and translates it
into a lower-level dataflow program on the right side that is guaranteed to be data-race- and
deadlock-free, while still allowing pipelining.  
\system uses the following five main passes: 
(1) capability type checking, to ensure that the source program correctly uses fences to mark conflicting memory accesses; 
(2) elaboration, to lower fences to explicit manipulation of affine memory permissions using the $\join(\dots)$ operator;
(3) control flow conversion, to convert sequential control flow to dataflow;
(4) linking, to modularly connect individually compiled functions into a single dataflow graph; 
and (5) operator selection and optimization, 
to lower the graph to an optimized, architecture-specific representation.
The soundness of the frontend passes (1) and (2) is 
certified with translation validation implemented in Rust, while
the core control flow conversion and linking passes (3) and (4)
are formally verified in Lean.
The last operator selection/optimization
pass is currently unverified, but can draw on orthogonal work 
in verifying graph rewrite passes~\cite{graphiti}.

\paragraphb{The Frontend: Capability Type Checking and Elaboration}
A programmer using \system writes their application in a sequential language, \lzero, which 
is implemented as a DSL embedded in Rust.
The goal of the frontend design is to give the core compiler hooks for 
safely determining when operations can proceed in parallel and when they must be sequentialized 
to remain sound, without the programmer reasoning about 
details of dataflow execution.  
To that end, we design: (1) a system of \emph{capability types} and \emph{fences} that 
ensure a program only type checks if potentially racing accesses are in separate fenced regions (\autoref{subsec:lzero-basics}), 
and (2) an elaboration pass that lowers fences into explicit permission tokens that 
the backend uses to provably preserve the typing guarantee (\autoref{subsec:ghost-lowering}).

The main idea of the capability types is that the user can type an array region $R$ as $\uniqtp$ for exclusive write access, or $\shrdtp$ for shared read access.
Crucially, these types are \emph{index-dependent}---the region $R$ need not be the entire array 
but instead a fine-grained slice, allowing the same array to be accessed in both exclusive and shared patterns.
This allows the compiler to identify operations on disjoint
array indices and prove that they can execute in parallel, enabling safe pipelining, a key goal of \system.
When the array access patterns are inherently sequential (e.g., sequential reads/writes to the same address), \system asks the programmer to insert \emph{fences} to
express ordering constraints explicitly. Once a program decorated with capabilities and fences type-checks, 
it is guaranteed to execute without data races, assuming the fences are respected. 

Since fences are not directly supported by the backend dataflow architecture, which 
enforces ordering through explicit control signals, we lower \lzero programs with fences
to an intermediate representation \lzeroghost, equipped with the explicit synchronization operator \join.
\system elaborates the abstract fences in \lzero programs into manipulation of \emph{ghost permission
variables} in \lzeroghost (\autoref{subsec:ghost-lowering}), reifying the fence's
ordering constraints as explicit data
dependencies between ghost variables, similar to control signals in the underlying architecture.
To ensure that the lowering process is sound, we adopt
a \emph{translation validation} approach, 
where we validate that permission tokens present in the lowered
\lzeroghost program reflect the original typing constraints and satisfy the soundness property assumed by \system's backend (\autoref{subsec:backend-assumptions}). 

\paragraphb{The Backend: Verified Control Flow Conversion and Linking}
The elaborated \lzeroghost program still uses sequential control flow.
As the next step, \system performs two formally verified core passes (the right half of \autoref{fig:pipeline})
to compile an \lzeroghost program to the dataflow calculus \lone.

The first is \emph{control flow conversion}
(\autoref{sec:compiler-cf}), which individually compiles each function in the elaborated \lzeroghost program 
to an \lone process by translating sequential control-flow
constructs (e.g., branching and tail-recursion) into pure dataflow operators (e.g., steer and carry in \autoref{fig:df-ops}).
While this pass is standard in dataflow compilers dating back to
TTDA~\cite{ttda}, it is particularly error-prone and challenging to
verify.
The core difficulty stems from the mismatched semantics of variables in a sequential program and channels in a dataflow program.
Unlike a sequential program, when a ``variable'' in dataflow is used in, e.g.,
both branches of a conditional, each use requires a \emph{fresh} communication
channel to receive the value.
Thus, to formally relate variables to channels, our simulation proofs need to carefully track unused
variables and the current path condition.
Additionally, for a dataflow graph to be safely invoked multiple times,
\system ensures that all unused values are
properly consumed at the end of execution to avoid deadlocks.

\begin{figure}[tb]
    \centering
    \includegraphics[width=\textwidth]{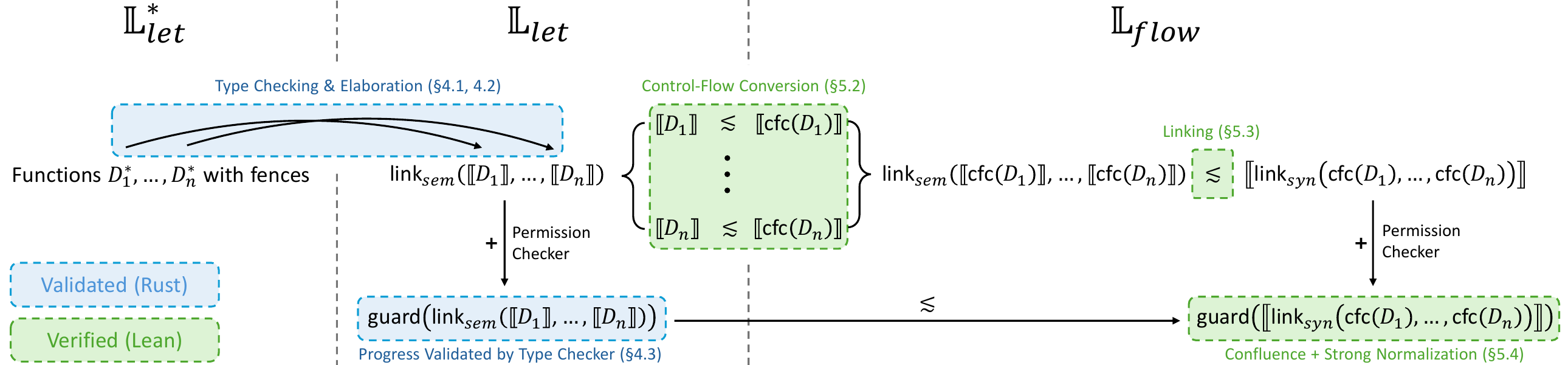}
    \caption{Proof structure overview of \system. $\lesssim$ denotes simulation; $\cfc$ is
    control flow conversion (\Cref{sec:compiler-cf}); $\linksem$ is the semantic linking of LTSs (\Cref{sec:compiler-linking}); $\linksyn$ is the syntactic linking of dataflow graphs (\Cref{sec:compiler-linking}); $\guard$ imposes
    permission checking on the base semantics (\Cref{sec:compiler-guarded}).}
    \Description{Proof structure overview.}
    \label{fig:proof-structure}
\end{figure}

The second verified core pass is \emph{linking} (\autoref{sec:compiler-linking}), which connects individually compiled dataflow graphs into a single graph representing the entire program.
Function calls in dataflow pose unique challenges for linking.
Unlike sequential programs with an explicit call stack, dataflow graphs distribute local variables across channel buffers that may hold values from multiple concurrent
function invocations.
To tackle these challenges, we place static restrictions (\autoref{subsec:backend-assumptions}) on \lzero to facilitate linking, and use a modular linking semantics inspired by interaction semantics~\cite{comp-compcert}.

To aid formal verification, we implement these core passes in Lean. This implementation is 
parametric in an abstract set of operators, making our compiler framework applicable to dataflow architectures with different instruction sets. 
We then prove the following final correctness results: (a) \emph{forward simulation}: the output dataflow graph has at least one deadlock-free schedule with the same 
behavior as the input sequential program; and (b) \emph{determinacy}: given a terminating \lzeroghost program 
that satisfies our type system's soundness property, the output \lone process is confluent and strongly normalizing, 
producing the same result regardless of execution schedule.

\autoref{fig:proof-structure} gives an overview of the proof structure of our compiler, showing how different guarantees 
of each component connect to establish our final correctness results. In particular,
we design the mechanized portion of the proofs to be modular in a few key ways to improve maintainability.
First, we separate the simulation proofs for control flow conversion and
linking, simplifying both and enabling modular replacement and verification
of functions with optimized dataflow graphs.
We then propagate our type system's soundness property as \emph{label restrictions} on the
labeled transition systems of \lzeroghost and \lone, and prove determinacy
independently from the core simulation proofs.

\paragraphb{Final Operator Selection and Optimizations}
At this point, we have constructed a dataflow graph in \lone that simulates the input \lzeroghost program and is determinate.
To interface with actual backend hardware (e.g., CGRAs~\cite{cgra-survey} or FPGAs), we lower certain pseudo-operators to architecture-specific operators (instruction selection), and perform unverified optimizations to improve the quality of the dataflow graph.
We define these transformations as term rewrites in \lone (\autoref{sec:compiler-opt}).
Combined with the memory ordering strategy in the type checker/elaborator and the core control-flow passes of the compiler,
we show in \autoref{sec:eval} that we can produce dataflow graphs with comparable quality to existing unverified dataflow compilers in RipTide~\cite{riptide} and LLVM CIRCT~\cite{circt}.

\section{Frontend: Capability Types and Fences}
\label{sec:typing}

In this section, we present the design of \system's frontend, which consists of a capability type system for our source language \lzero (\autoref{subsec:lzero-basics}), an elaboration pass to an intermediate language \lzeroghost (\autoref{subsec:ghost-lowering}) with a translation validation phase to ensure that the produced \lzeroghost program satisfies the soundness property assumed by later verified compiler passes (\autoref{subsec:backend-assumptions}).
While both \lzero and \lzeroghost are sequential languages and do not need memory synchronization by themselves, the goal of the type system and elaboration is to soundly make \emph{implicit} memory dependencies \emph{explicit}, so that we can guarantee the determinacy of the dataflow graph produced by the core verified passes.

\begin{figure}[tb]
\centering
\small
\(
\begin{array}{llcl@{\quad\,\,}llcl}
\textit{Op} & \mathsf{op} & \bnfdef & \mathbf{add} \bnfalt \mathbf{sub} \bnfalt \mathbf{mul} \bnfalt \mathbf{ld} \bnfalt \mathbf{st} \bnfalt \cdots 
&
\textit{Value} & v & \bnfdef & n \bnfalt \mathbf{true} \bnfalt \mathbf{false} \bnfalt \mathbf{()} 
\\
\textit{Expr} & E & \bnfdef & \ret(\vv{x}) \bnfalt \tailcall(\vv{x}) \bnfalt \mathbf{let}~\vv{y} = \mathsf{op}(\vv{x});~E
&
\textit{Type} & \tau & \bnfdef & \mathbf{bool} \bnfalt \mathbf{unit} \bnfalt [\tau;~\texttt{N}]
\\ 
& & & \bnfalt \mathbf{let}~\vv{y} = f(\vv{x});~E \bnfalt \mathbf{let}~y = v;~E  
& &  &  & \bnfalt \mathbf{u8} \bnfalt \mathbf{u16} \bnfalt \mathbf{u32}  \bnfalt \cdots
\\
& & & \bnfalt \mathbf{if}~x~\mathbf{then}~E_1~\mathbf{else}~E_2 \bnfalt \fence~E
&
\textit{Cap.} & \capsym & \bnfdef & \mathsf{uniq@}R \bnfalt \mathsf{shrd@}R \bnfalt \capsym_{1} \capsep \capsym_{2}
\\
\textit{Defn} & D & \bnfdef  & \mathbf{def}~f(\vv{x: \tau}) \langle~\vv{A : \capsym}~\rangle \to \vv{\tau_o} = E 
&
\textit{R-expr} & r & \bnfdef & n \bnfalt x \bnfalt r + r \bnfalt r - r \bnfalt r * r 
\\
\textit{Program} & P & \bnfdef & \vv{D} 
&
\textit{Region} & R & \bnfdef & r..r 
\\
\end{array}
\)
\caption{Syntax of \lzero.}
\Description{Syntax of \lzero.}
\label{fig:syntax-l0}
\end{figure}

\subsection{\texorpdfstring{\lzero}{L*let} with Capability Types and Fences}
\label{subsec:lzero-basics}

This section defines \lzero,
which is a first-order sequential language with capability types and a construct called a \emph{fence} to explicitly annotate memory dependencies.

\paragraphb{Syntax}
\autoref{fig:syntax-l0} shows the syntax of \lzero.
A top-level \lzero program $P$ is made up of a list of function definitions,
and the body of each function is an expression consisting of 
$\m{let}$ bindings, conditional expressions, returns, and tail calls.
A key feature of \lzero is that every expression can be optionally preceded by
a fence (\fence), intuitively denoting memory dependencies from operations after the fence to operations before it.
To help the type system infer suitable capabilities,
each \lzero function is annotated with two lists of typed parameters: (1) ordinary variables $\vv{x: \tau}$, and (2) capability-annotated array inputs $\vv{A : C}$ (assumed to be disjoint from $\vv{x}$).
A capability type $C$ describes allowed access permissions to a precise \emph{region} of the annotated array.
Syntactically, this permission can be shared \(\shrd{R}\) or unique \(\uniq{R}\),
where $R$ is a range $[r_1..r_2)$ of indices.

\begin{figure}[htb]
\centering
\small
\begin{mathpar}

\inferrule[Load]{
  \env(i) = \mathbf{int}
  \quad
  \checkcolor{
  \cond \vDash 
  \shrd{i} \le \capa[A]
  }
  \\\\
  \judg{\fnctx ; \env[y \mapsto \mathbf{int}]}{\capa[A \mapsto \capa[A] \setminus \shrd{i}]}{f, \cond}{E}{\tau}
}{
  \judg{\fnctx ; \env}{\capa}{f, \cond}{\mathbf{let}~y~\mathbf{=}~\mathbf{ld}(A,i);~E}{\tau}
}

\inferrule[Fence]{
  \fnctx(f) = \mathbf{def}~f(\vv{x_i: \tau_i}) \langle~\vv{A : \capsym}~\rangle \to \vv{\tau_o}
  \\\\
  \judg{\fnctx ; \env}{\vv{A: \capsym}}{f, \cond}{E}{\tau}
}{
  \judg{\fnctx ; \env}{\capa}{f, \cond}{\fence~E}{\tau}
}

\inferrule[Store]{
  \env(i) = \mathbf{int}
  \quad
  \env \vdash v : \mathbf{int}
  \quad
  \checkcolor{
  \cond \vDash
  \uniq{i} \le \capa[A]
  }
  \\\\
  \judg{\fnctx ; \env}{\capa[A \mapsto \capa[A] \setminus \uniq{i}]}{f, \cond}{E}{\tau}
}{
  \judg{\fnctx ; \env}{\capa}{f, \cond}{\mathbf{let}~()~\mathbf{=}~\mathbf{st}(A,i,v);~E}{\tau}
}

\inferrule[Tail-Call]{
  \fnctx(f) = \mathbf{def}~f(\vv{x_i: \tau_i}) \langle~\vv{A : \capsym}~\rangle \to \vv{\tau_o}
  \\\\
  \env(\vi) = \vv{\tau_i}
  \quad
  \checkcolor{
  \cond \vDash \forall a \in \vv{A}.
  \,\capsym[\vv{i/x_i}] \le \capa[a]\,
  }
}{
  \judg{\fnctx ; \env}{\capa}{f, \cond}{\tailcall(\vi)}{\vv{\tau_o}}
}

\end{mathpar}
\caption{Selected typing rules for \lzero.}
\Description{Selected typing rules for \lzero.}
\label{fig:key-typing-rules}
\end{figure}

\paragraphb{Capability Type System}
We equip \lzero with a capability type system to ensure that user-provided capability annotations correctly describe the program's actual memory access patterns and that the fences in the program soundly overapproximate all memory dependencies.
Information inferred from this type system facilitates the translation of \lzero to \lzeroghost (\autoref{subsec:ghost-lowering}), which has a more explicit representation of capabilities and is more compatible with later compilation passes.
There are two key challenges that our type system must address to be precise enough for the purpose of pipelining (\autoref{sec:background}):
(1) it needs to distinguish between read and write accesses since reads can be done safely in parallel, but writes (to the same location) need to be synchronized; and
(2) it needs to distinguish accesses to different regions of an array to avoid over-synchronization.
As a result, each capability $C$ has two components: a $\mathsf{shrd}$/$\mathsf{uniq}$ component for distinguishing read/write accesses, and a region $R$ component for specifying the precise indices.

We now sketch the design of our type system, where the main typing judgment has the form $\judg{\fnctx ; \env}{\capa}{f, \cond}{E}{\tau}$.
Here, $f$ denotes the current function, $\fnctx$ maps function names to their signatures, 
$\env$ is the typing context for ordinary variables, $E$ is the expression being checked, and $\tau$ is its final type.
Additionally, two important contexts that we need to include in the type judgment are 
the capability context $\capa$, which is a map from array names to capabilities $C$, and the propositional context $\cond$, which is a collection of logical facts from branches or let bindings that overapproximate the state of ordinary variables.
In particular, both $\capa$ and $\cond$ can mention ordinary variables in $\env$.

\begin{figure}[t]
  \centering 
  \includegraphics[width=\textwidth]{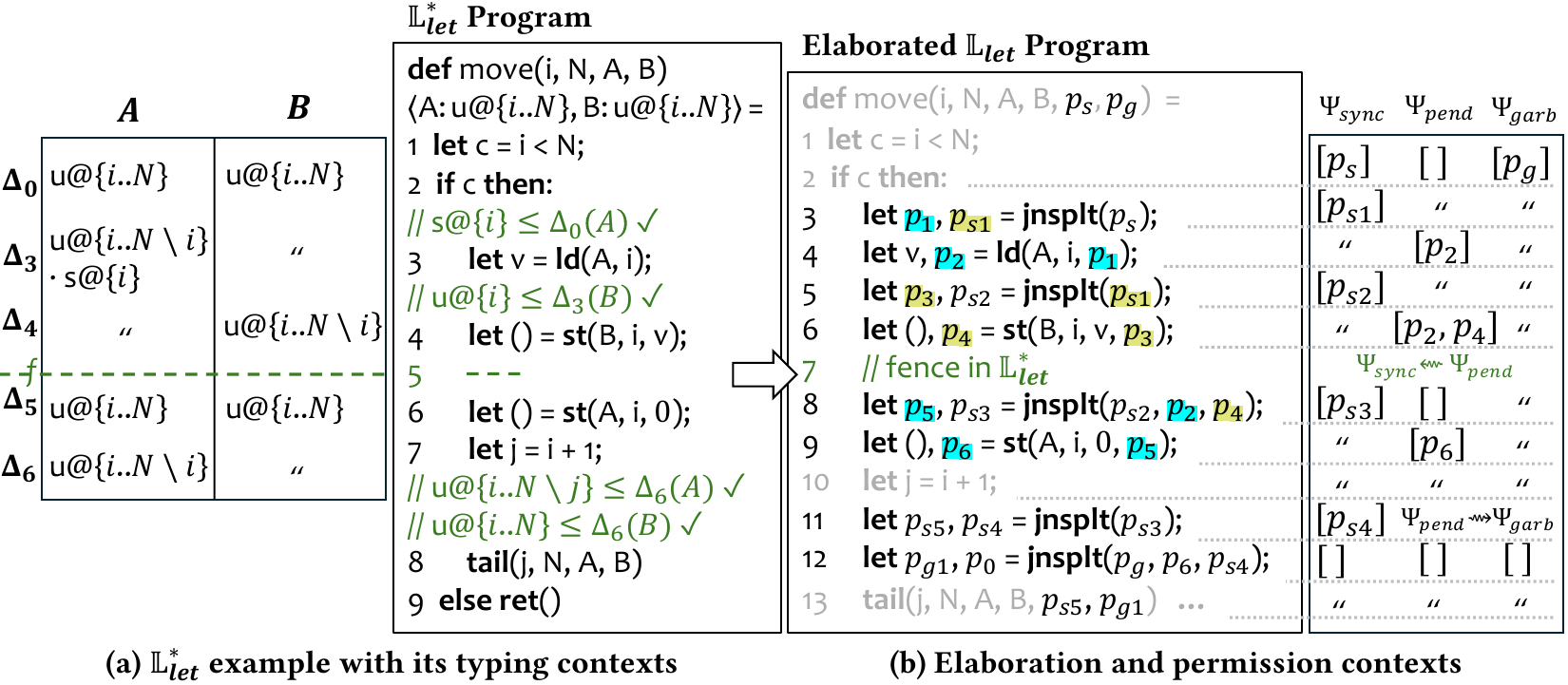}
  \caption{An example program \Lzerocode{move} in \lzero, along with its typing contexts, elaborated \lzeroghost program, and the permission contexts during elaboration. Permissions for arrays \aperm{$A$} and \bperm{$B$} are highlighted. $\mathsf{u}@\{\dots\}$ and $\mathsf{s}@\{\dots\}$ are abbreviations for $\uniq{\dots}$ and $\shrd{\dots}$, respectively.}
  \Description{Typing example.}
  \label{fig:move-typing}
\end{figure}

\autoref{fig:key-typing-rules} presents selected key rules, and the full formulation can be found in our appendix~\cite{appendix}.
To explain the intuition behind these rules, we use the concrete example in \autoref{fig:move-typing}~(a), which is an \lzero function that copies the contents of array \textsf{A} to \textsf{B}, and then clears \textsf{A} to zero.
When loading from or storing to an array, we need to check that the current capability context $\capa$ contains suitable permissions for accessing the target array ($\mathsf{shrd}$ for load and $\mathsf{uniq}$ for store) at the specific index being accessed.
These checks are described by the \textsc{Load} and \textsc{Store} rules in \autoref{fig:key-typing-rules}.
This is not a purely syntactic check since our capabilities are dependent on ordinary variables (e.g., to track indices), and as a result, premises such as $\checkcolor{\cond \vDash \shrd{i} \le \capa[A]}$ involve SMT queries to check whether we have suitable capabilities under the current propositional context $\cond$ (where we order $\mathsf{shrd} \leq \mathsf{uniq}$ and the ranges are ordered by set inclusion).
These SMT queries are represented in green comments in our example in \autoref{fig:move-typing}~(a).
After the load (resp. store), the suitable permission for the read (resp. written) index is consumed and removed from the capability context $\capa$ when type checking the rest of the expression.
One subtlety in removing capabilities is that, to allow parallel loads, we have $\mathsf{shrd} \setminus \mathsf{shrd} = \mathsf{shrd}$ and $\mathsf{uniq} \setminus \mathsf{shrd} = \mathsf{shrd}$.
In our example, consider the load on line 3.
Before the load, we have $\capa_0[A] = \uniq{i .. N}$, and after the load, we have $\capa_3[A] = \uniq{i .. N \setminus i} \capsep \shrd{i}$, where $\capsep$ denotes the union of capabilities.

In order to write to $i$ again on line 6, the \lzero program must insert a fence to indicate a memory dependency.
A fenced expression $\fence~E$ in \lzero intuitively ``resets'' the capability context to the initial function context, as formally described by rule \textsc{Fence}.
In our example, the fence on line 5 resets the next capability context to $\capa_5 = \capa_0$ before type checking the rest.

Finally, when calling another function or doing a tail call, the \textsc{Tail-Call} rule is used to check that suitable capabilities required by the callee function are present in the current capability context.

\subsection{Ghost Permissions Inference and Fence-Removal}\label{subsec:ghost-lowering}

The language \lzero only requires light annotations to soundly overapproximate memory dependencies in the program, but dataflow architectures do not have a fence primitive.
Instead, each $\m{LD}$ and $\m{ST}$ operator takes an additional argument 
(a control signal) that indicates when the operation can safely proceed (\Cref{sec:background}).
Therefore, to bridge this gap, \system lowers \lzero to an intermediate language called \lzeroghost with more explicit forms of memory dependencies.

We demonstrate the lowering using the same example in \autoref{fig:move-typing}~(b), which is the lowered \lzeroghost program from \autoref{fig:move-typing}~(a).
There are several differences in this intermediate representation (highlighted in the figure).
First, \lzeroghost removes all capability types and fences, but extends \lzero with two ``ghost'' constructs:
ghost variables representing affine permission tokens and the ghost
operator \join, which is used to \textbf{j}oi\textbf{n} input
permission tokens and then \textbf{spl}i\textbf{t} them in a suitable way.
Semantically, $\mathbf{let}~p',~p'' = \join(p_1, \ldots, p_n)$ 
combines $n$ permission variables and splits the result into two fresh
permissions, representing the needed permission for the operation immediately after this \join and the remaining permissions for subsequent operations.
Another main difference is that operators and functions now accept and return explicit permission variables.
For example, memory operations now consume and produce one additional permission variable, and functions accept two additional permission arguments $p_{\mathit{sync}}$ and $p_{\mathit{garb}}$ (abbreviated as $p_s$ and $p_g$, respectively, in \autoref{fig:move-typing}~(b)).

While the detailed lowering definition can be found in our appendix~\cite{appendix}, we highlight some key aspects.
To guide the lowering process, we maintain a permission context
$\Psi = (\Psi_{\mathit{sync}}, \Psi_{\mathit{pend}}, \Psi_{\mathit{garb}})$.
$\Psi_{\mathit{sync}}$ is initialized with $p_{\mathit{sync}}$, which holds
synchronized permissions available for immediate use---they are inherited from
capability types, and are therefore \emph{guaranteed sufficient} up to the next fence point
(or the end of the function if there is no fence).
$\Psi_{\mathit{garb}}$ is initialized with $p_{\mathit{garb}}$, which is needed
to ``garbage collect'' permissions that are no longer needed and can be
returned to the caller, enabling permission reuse in subsequent operations. Finally,
$\Psi_{\mathit{pend}}$ tracks permissions from issued operations awaiting
synchronization (moving to $\Psi_{\mathit{sync}}$ at fences) or garbage collection (moving to $\Psi_{\mathit{garb}}$ at tail calls and returns). Together, these contexts let the lowering soundly materialize the capabilities and fences from \lzero while avoiding unnecessary memory dependencies in the resulting \lzeroghost program, which is crucial for safe pipelining in the final \lone program.

To see how these contexts are used and how synthesized permissions reflect the original typing constraints,
consider lines 3-9 in \autoref{fig:move-typing} (b).
When translating a $\m{ld}$/$\m{st}$, we synthesize a \join that splits
$\Psi_{\mathit{sync}}$ into a permission for the $\m{ld}$/$\m{st}$ and remaining permissions. The returned permission from the $\m{ld}$/$\m{st}$ moves into $\Psi_{\mathit{pend}}$, and the lowering process continues
with the remaining permissions in $\Psi_{\mathit{sync}}$ until the next fence, which shifts all pending permissions back into $\Psi_{\mathit{sync}}$.
Operationally, these permission variables enforce sound memory synchronization.
For example, the $\m{ld}$ on line 4 and the $\m{st}$ on line 6 depend on \emph{disjoint} permissions and can be executed in parallel, while the $\m{st}$ on line 9 needs to wait for the $\m{ld}$ on line 4 to produce the permission \aperm{$p_2$}.
Function calls are similar to $\m{ld}$/$\m{st}$ except that we synthesize \emph{two} $\join$s
for $p_{\mathit{sync}}$ and $p_{\mathit{garb}}$, respectively
(lines 11-12).

\subsection{Backend Assumptions}
\label{subsec:backend-assumptions}

The capability type checking (\autoref{subsec:lzero-basics}) and fence-removal (\autoref{subsec:ghost-lowering}) passes of our frontend ensure that implicit memory dependencies (implied by the control flow in \lzero) are explicitly converted to data dependencies on ghost permission variables.
Since these passes are currently not formally verified in Lean, we discuss two important assumptions that our verified backend (\autoref{sec:compiler}) places on the frontend to ensure compilation correctness, as well as how they are currently validated.

\paragraphb{Soundness of Ghost Permissions}
While the ghost permission lowering pass intuitively materializes the
memory access guarantees established by \lzero's type system (\autoref{subsec:lzero-basics}), how do we know that the synthesized permissions indeed preclude any conflicting memory accesses once further lowered to \lone?
For example, a buggy lowering pass might ignore capabilities entirely and synthesize a $\join$ that produces more permissions than it consumes.
This would allow parallel writes to the same memory location, introducing data races in the final \lone program.
To address this, our frontend uses \emph{translation validation} to certify the soundness of the inferred permission tokens, which our verified backend ultimately relies on to ensure determinacy (\autoref{sec:compiler-guarded}).
Note that the translation validator itself is not formally verified, and it serves as an additional safeguard for the frontend.

To formalize the concept of permissions, we rely on \emph{partial commutative monoids} (PCMs)~\cite{iris}, which provide a mathematical framework for reasoning about resource sharing and disjointness.
For our frontend, we use the PCM instance of \emph{fractional permissions}~\cite{fractional-perm}: resources are associated with a fraction $f \in [0, 1]$, where $f = 1$ means exclusive access, and any fraction $0 < f < 1$ represents shared access.
We define dependent permission specifications for each operator and function in \lzeroghost, each of which includes a precondition $\precond$ (mapping input values to required permissions) and a postcondition $\postcond$ (mapping input and output values to returned permissions). For example, the store operator $\m{st}$ has
$\precond_{\m{st}}(A, i) = \m{1@A\{i\}}$ and
$\postcond_{\m{st}}(A, i, v) = \m{1@A\{i\}}$ (i.e., exclusive access to $A[i]$). For \join, we specify that the PCM sum of the input and output permissions must be equal, ensuring that it does not create or destroy permissions.

The \emph{guarded semantics} later defined in \autoref{sec:compiler-guarded} dynamically checks that the permission specification above is satisfied and gets stuck otherwise.
Hence, the exact soundness property validated in this pass (assumed later in \autoref{thm:strong-norm}) is the progress of the \lzeroghost program under the guarded semantics.

\paragraphb{Static Restrictions on \lzero}
The dataflow model that we target in this work uses FIFO queues as channels and requires a static dataflow graph topology.
These architectural constraints complicate the compilation of function calls: when a function has multiple (parallel) callers, the compiler must merge their input arguments and route return values to the correct caller.
A static topology requires these routing decisions to be fixed ahead of time, while FIFO channels make arbitrary interleavings of parallel calls difficult because operators must process their inputs in order.

While there are multiple (unverified) solutions to this challenge 
(\autoref{sec:related}), we adopt an approach similar to RipTide~\cite{riptide}, where the control flow of \lzeroghost is assumed to satisfy two static restrictions: (1) a function can be statically called at most once outside its definition, but multiple tail-recursive calls within the function definition are allowed; and (2) no mutual recursion is allowed.
These restrictions are validated by the backend when parsing the input \lzeroghost program from the frontend.

Requirement (1) allows a function to tail-call itself multiple times, such as
$\mathbf{def}~f(x) = \mathbf{if}~x \le 1~\mathbf{then}~\ret(x)~\mathbf{else}~(\mathbf{if}~x\%2 = 0~\mathbf{then}~\tailcall(x / 2)~\mathbf{else}~\tailcall(3x + 1))$.
However, \emph{outside} its own definition, there can be at most one static call site to $f$.
Requirement (2) forbids mutual recursion.
In most dataflow architectures, FIFO channels have bounded buffers, and they block the sender if they are full (i.e., backpressure).
As a result, non-tail mutual recursion such as
$\mathbf{def}~f(x) = g(x) + 1; \mathbf{def}~g(x) = f(x) + 1$
causes a \emph{deadlock}, because it induces an unbounded ``stack.''

These restrictions do not limit the expressiveness of \lzeroghost and \lzero, since any structured control flow (i.e., with branching and loops) can be transformed to satisfy these restrictions: branching is built-in, and loops can be transformed to tail-recursive calls.

They do, however, reduce the ergonomics of writing \lzeroghost directly, because users must currently encode control flow in this normalized form.
In future work, we plan to automate this process with desugaring/normalization passes, or to use \lzeroghost as an IR for dataflow compilation.

\begin{figure}[tb]
    \small
    \[
    \begin{array}{c}
    ((\mathbf{def}~f(\vv{y}) = E), \mathsf{init}, \sigma) \xrightarrow{\inputl{\vv{v}}} ((\mathbf{def}~f(\vv{y}) = E), E, \{ \vv{y} \mapsto \vv{v} \}) \\[6pt]
    
    (D, \mathbf{ret}(\vv{x}), \sigma) \xrightarrow{\outputl{\sigma(\vv{x})}} (D, \mathsf{init}, \emptyset) \\[6pt]

    ((\mathbf{def}~f(\vv{y}) = E), \mathbf{tail}(\vv{x}), \sigma) \xrightarrow{\tau} ((\mathbf{def}~f(\vv{y}) = E), E, \{ \vv{y} \mapsto \sigma(\vv{x}) \}) \\[6pt]

    (D, (\mathbf{let}~\vy = \mathsf{op}(\vx);~E), \sigma) \xrightarrow{\yieldl{\mathsf{op}, \sigma(\vx), \vv{v}}} (D, E, \sigma[\vy \mapsto \vv{v}]) \\[6pt]

    (D, \mathbf{if}~x~\mathbf{then}~E_\mathbf{true}~\mathbf{else}~E_\mathbf{false}, \sigma) \xrightarrow{\tau}
        (D, E_{\sigma(x)}, \sigma) \quad \text{if } \sigma(x) \in \{ \mathbf{false}, \mathbf{true} \}
    \end{array}
    \]
    \caption[]{Small-step operational semantics of an \lzeroghost definition $D$, denoted $\interp{D} := ((D, \mathsf{init}, \emptyset), \to) \in \lts{\configlzero}{\labelset_O}$, with the set of configurations $\configlzero := \mathit{Defn} \times (\mathit{Expr} \sqcup \{ \mathsf{init} \}) \times (\mathit{Variable} \rightharpoonup \mathit{Value})$. For a list of variables $\vx = x_1, \dots, x_n$, we use $\sigma(\vx)$ to denote the list of values $\sigma(x_1), \dots, \sigma(x_n)$, assuming they are all defined.}
    \Description{Small-step operational semantics of \lzeroghost.}
    \label{fig:semantics-l0}
\end{figure}

\begin{figure}[tb]
    \small
    \vspace{-8pt}
\[
\begin{array}{c}

\multicolumn{1}{c}{
\begin{array}{rcll}
\textit{Proc} & \pi & ::= \mathsf{op}(\vv{c}) \to \vv{d} & \mathsf{op} \in O \\
            & & \; \; \ \mid \mathsf{merge}_\mathit{s}(c_1, \vv{c_2}, \vv{c_3}) \to \vv{d} & {s \in \{\mathsf{init}, \mathsf{left}, \mathsf{right}\}} \\
            & & \; \; \ \mid \mathsf{steer}_f(c_1, \vv{c_2}) \to \vv{d} & {f \in \{ \bot, \top \}} \\
            & & \; \; \ \mid \mathsf{fork}(c) \to \vv{d} \mid \mathsf{order}(\vv{c}) \to d \mid \mathsf{sink}(\vv{c}) \\
            & & \; \; \ \mid \mathsf{forward}(\vv{c}) \to \vv{d} \mid \mathsf{const}_v(c) \to \vv{d} \\
            & & \; \; \ \mid \pi_1 \para \pi_2 \mid \nu c.\ \pi
\end{array}
} \\

\\
(\pi, \sigma) \xrightarrow{\inputl{\vv{v}}} (\pi, \push(\vv{c_{in}}, \vv{v}, \sigma)) \qquad

\inferrule*{
    \pop(\vv{c_{out}}, \sigma) = (\vv{v}, \sigma')
}{
    (\pi, \sigma) \xrightarrow{\outputl{\vv{v}}} (\pi, \sigma')
} \qquad

\inferrule*{
    (\pi_1, \sigma) \xrightarrow{l} (\pi'_1, \sigma')
}{
    (\pi_1 \para \pi_2, \sigma)
    \xrightarrow{l}
    (\pi_1' \para \pi_2, \sigma')
}
\\[12pt]

\inferrule*[right=Sync-Op]{
    \pop(\vv{c}, \sigma) = (\vv{v_{in}}, \sigma')
}{
    (\apop(\vv{c}) \to \vv{d}, \sigma)
    \xrightarrow{\yieldl{\apop, \vv{v_{in}}, \vv{v_{out}}}}
    (\apop(\vv{c}) \to \vv{d}, \push(\vv{d}, \vv{v_{out}}, \sigma'))
}



\end{array}
\]
    \caption[]{Syntax and selected rules of the semantics of \lone processes. We denote the semantics of a process $\pi$ as $\interp{\pi}_{\vv{c_{in}}, \vv{c_{out}}} := ((\pi, \emptyset), \to) \in \lts{\configlone}{\labelset_O}$, with the set of configurations $\configlone := \mathit{Proc} \times (\textit{Channel Name} \rightharpoonup \mathit{Value}^*)$. Here, $\vv{c_{in}}$ and $\vv{c_{out}}$ are distinguished lists of input and output channels. $\push$ and $\pop$ are auxiliary functions for pushing and popping values from channel buffers.}
    \Description{Small-step operational semantics of \lone.}
    \label{fig:semantics-l1}
\end{figure}

\section{Compiler Design and Verification}
\label{sec:compiler}

In this section, we detail the design and formal verification of the core compilation passes of \system, covering the verified portions of our proof structure overview in \autoref{fig:proof-structure}.

In \autoref{sec:compiler-basic}, we introduce our dataflow calculus \lone and the semantics of \lzeroghost. 
In \autoref{sec:compiler-cf} and \autoref{sec:compiler-linking}, we formally define the two core passes of control flow conversion (\cfc) and linking (\linksem/\linksyn), and sketch their forward simulation proofs.
In \autoref{sec:compiler-guarded}, we define the guarded semantics (\guard) and prove the determinacy of the final dataflow graph.
Finally, in \autoref{sec:compiler-opt}, we briefly cover our optimization passes as graph rewrites on \lone processes.

\subsection{Dataflow Calculus \texorpdfstring{\lone}{Lflow} and the Semantics of \texorpdfstring{\lzeroghost}{Llet}}
\label{sec:compiler-basic}

We first introduce a simple calculus \lone as our target language.
Its syntax can be found in \autoref{fig:semantics-l1}.
In essence, an \lone process is a parallel composition of \emph{atomic processes}, where each atomic process is a recursive process that waits for inputs, performs computation, and produces some outputs.

We classify atomic processes into two categories: \emph{asynchronous} and \emph{synchronous} operators.
An operator is asynchronous if its input/output behavior may depend on its input values and local state (e.g., steer in \autoref{sec:background}), whereas a synchronous operator reads and then produces exactly one value from/to each input/output channel (e.g., arithmetic or memory operators).

Asynchronous operators such as \apsteer and \apmerge are built-in, because our core passes depend on their semantics to convert control flow and perform optimizations, and we will discuss their semantics shortly.
Besides asynchronous operators, the syntax and semantics of \lone are parametric in a customizable set $O$ of synchronous operators (previously specialized to $\mathit{Op}$ in \autoref{fig:syntax-l0}).
These operators do not need special treatment in compilation due to their more uniform behavior.
This separation of built-in asynchronous operators and custom synchronous operators allows \system to easily target dataflow architectures with different instruction sets.
Further, the similar interface of a synchronous operator and a function in \lzeroghost also allows \system to treat function calls modularly as operator calls and perform linking in a separate pass.

Finally, atomic processes in \lone can be used in parallel composition $\pi_1 \para \pi_2$ and channel restriction $\nu c.~\pi$, with the usual structural congruence rules of commutativity and associativity of $\para$, scope extrusion, and alpha renaming.

\paragraphb{Semantics of \lzeroghost and \lone}
We use labeled transition systems (LTSs) to model the semantics of both \lzeroghost and \lone.
As notation, we use $\lts{C}{L}$ as a shorthand for the set of relations $R \subseteq C \times L \times C$, with states $C$ and labels $L$, and a distinguished initial state $c_0 \in C$.
For any $(c_0, \to) \in \lts{C}{L}$, we use $c \xrightarrow{\mt{ls}} c'$ to denote zero or more steps from $c$ to $c'$ with some string $\mt{ls}$ of labels. 

The small-step operational semantics of \lzeroghost definitions and \lone processes are shown in \autoref{fig:semantics-l0} and \autoref{fig:semantics-l1}.
Both LTSs share four types of labels:
\[
    \labelset_O \ni l ::=
        \tau \mid
        \inputl{\vv{v}} \mid
        \outputl{\vv{v}} \mid
        \yieldl{\mathsf{op}, \vv{v_{in}}, \vv{v_{out}}} \quad \mathsf{op} \in \mathit{O}
\]

The silent label $\tau$ denotes an internal step without external effects;
an $\inputl{\vv{v}}$ label provides initial input $\vv{v}$
to the program/process; an $\outputl{\vv{v}}$ label denotes final output $\vv{v}$
produced by the program/process; and finally,
a $\yieldl{\mathsf{op}, \vv{v_{in}}, \vv{v_{out}}}$ label denotes calling
either an operator or function with inputs $\vv{v_{in}}$,
and transitioning to a state assuming the outputs to be $\vv{v_{out}}$.

The transitions of \lzeroghost are mostly straightforward.
Initially, a function accepts an input label and initializes the parameters.
A return expression produces an output label with the return values.
A tail call produces a silent label and re-initializes the current function with tail call arguments.
Note, however, that function and operator calls are left uninterpreted in the \lzeroghost semantics: they simply generate yield labels and non-deterministically transition to a state assuming the return values of the call.
This is intended to make our proofs more modular, as we can separately prove forward simulations of individual function compilation (\autoref{sec:compiler-cf}) and linking (\autoref{sec:compiler-linking}).
This is also similar to uninterpreted events in interaction trees~\cite{itrees}.

On the \lone side (\autoref{fig:semantics-l1}), we show selected rules for inputs/outputs, parallel composition, and synchronous operators, and describe the semantics of built-in operators in the text below.
The $\apsteer_f(c_1, \vv{c_2}) \to \vv{d}$ operator reads a Boolean value from the ``decider'' channel $c_1$ and input values from $\vv{c_2}$. If the decider equals the filter $f$, the input values are forwarded to $\vv{d}$; otherwise, the input values are discarded.
The $\apmerge_s(c_1, \vv{c_2}, \vv{c_3}) \to \vv{d}$ operator has three states.
At the initial state $s = \textsf{init}$, it reads the decider value from $c_1$.
If the decider is true, it transitions to $\textsf{left}$, and otherwise it transitions to $\textsf{right}$.
Then, at state $\textsf{left}$ (resp. $\textsf{right}$), it forwards the input from $\vv{c_2}$ (resp. $\vv{c_3}$) to $\vv{d}$, and then transitions back to $\textsf{init}$.
The carry operator in \autoref{fig:df-ops} for representing loop variables is exactly $\apmerge_\textsf{left}$.
The $\apfork$ operator duplicates its input to multiple outputs.
The $\aporder$ operator is a low-level synchronization primitive that waits for all its inputs to be ready and then forwards the first input.
The $\apsink$ operator consumes its input without any output.
The $\apforward$ operator simply forwards its input to its output.
Finally, $\apconst_v(c)$ pushes the constant $v$ to the output whenever it receives any value from the input $c$.

To conclude our semantics setup, we define a notion of simulation $\iosim$ 
for LTSs with label set $\labelset_O$. Intuitively, this says that if the LTS $(c_0, \to)$ can perform one step, then a related state in $(c'_0, \to')$ can also make multiple steps to reach another related state, with some restrictions such as no silent steps before an input label.
This definition is stricter than the usual notion of weak simulation~\cite{intro-bisim}, due to technical reasons in the simulation proofs of linking.

\begin{definition}[IO-Restricted Weak Simulation]
    Given two transition systems $(c_0, \to) \in \lts{C}{\labelset_O}$ and $(c'_0, \to') \in \lts{C'}{\labelset_O}$, a relation $R \subseteq C \times C'$ is an \emph{IO-restricted weak simulation} if $(c_0, c'_0) \in R$; and for every $(c, c') \in R$ and $l \in \labelset_O$ with $c \xrightarrow{l} c_1$, there exists $c'_1$ with $(c_1, c'_1) \in R$ such that: (1) if $l = \tau$, then $c' \xrightarrow{\tau^*} c'_1$; (2) if $l = \inputl{\vv{v}}$, then $c' \xrightarrow{l \tau^*} c'_1$; (3) if $l = \outputl{\vv{v}}$, then $c' \xrightarrow{\tau^* l} c'_1$; (4) if $l = \yieldl{\mathsf{op}, \vv{v_{in}}, \vv{v_{out}}}$, then $c' \xrightarrow{l} c'_1$.
    If there exists such a relation $R$, we say that $(c_0, \to) \iosim (c'_0, \to')$.
\end{definition}

\subsection{Control Flow Conversion}
\label{sec:compiler-cf}

In this section, we describe the core control flow conversion (CFC) pass of \system,
which, in essence, eliminates all control flow in the input \lzeroghost function,
and produces an \lone process with only dataflow.
We first define the compilation of each \lzeroghost construct,
and then sketch a proof of its forward simulation.
Throughout, we assume that we are compiling a single \lzeroghost function, parametric in the operator set $O$ (which can contain base operators or other \lzeroghost function symbols).

\paragraphb{Compiling a Function Definition}
The compiled \lone process of an \lzeroghost definition $D$, denoted $\cfc(D, \vv{c}, \vv{d})$, is defined below, where $\vv{c}$ (resp. $\vv{d}$) are free channels in $\cfc(D, \vv{c}, \vv{d})$ to represent input (resp. output) channels of the entire process, and bound names in $\nu$ are distinct and fresh.
\begin{align*}
\cfc(\mathbf{def}~f(\vx) = E, \vv{c}, \vv{d}) :=\ &\nu~\vv{c}', \vv{d}', t, \vv{r}, \vv{r}'.~
    \apmerge_\mathsf{left}(t, \vv{c}, \vv{r}) \to \vv{c}' \para & \text{\small (Merge inputs/tail call)} \\
    &\cfc(E, \{ \vx \mapsto \vv{c}' \}, t, \vv{d}', \vv{r}') \para &\text{\small (Compile body)} \\
    &\apsteer_\bot(t, \vv{d}') \to \vv{d} \para \apsteer_\top(t, \vv{r}') \to \vv{r} & \text{\small (Route outputs)}
\end{align*}

The definition consists of three parts, where the first and last lines are for initialization and cleanup.
The compilation of the function body $E$ in the middle is denoted $\cfc(E, \Sigma, t, \vv{d}, \vv{r})$, where
$\Sigma$ is a map from unused \lzeroghost variables to channel names; 
$t$ is a channel name used for sending a Boolean flag indicating whether the expression ends with a return or a tail call, which we will call the \emph{tail condition} below; $\vv{d}$ are channels used for sending return values; and $\vv{r}$ are channels used for sending tail call arguments.

The trickiness of compilation comes from determining the correct input and output channel 
routing, which depends on the different ways a function is called and terminates.
Recall that an \lzeroghost function can receive (non-recursive) external calls or tail-recursive calls within the function body (a design choice due to the difficulty with dataflow function calls discussed in \autoref{sec:overview}).
To produce a dataflow graph that handles both cases, we need a \apmerge operator at the beginning to combine call arguments from both situations.
The initial state of the \apmerge is set to $\mathsf{left}$, so that initially, the function will directly forward the input arguments to the function body, without requiring a decider value.
In turn, to handle outputs, \system must suitably route the output channels $\vv{d}'$ and $\vv{r}'$ based on the tail condition $t$: if $t$ is true (i.e., tail call), then the tail arguments $\vv{r}'$ are routed to the initial merge for recursion and $\vv{d}'$ are discarded via a \apsteer; if $t$ is false (i.e., return), then $\vv{d}'$ are routed to the final outputs $\vv{d}$, and $\vv{r}'$ are discarded by the second \apsteer.

\paragraphb{Compiling an Operator Call}
Handling operator (or function) calls is comparatively simple; an \lzeroghost operator is mapped to the corresponding atomic process in \lone, and suitable output channels are added to the context, where $\Sigma(\vx) \equiv \Sigma(x_1), \dots, \Sigma(x_n)$ (assuming all defined) for $\vx \equiv x_1, \dots, x_n$.
\begin{gather*}
    \cfc((\mathbf{let}~\vy = \apop(\vx);~E), \Sigma, t, \vv{d}, \vv{r}) :=
        \nu \vv{d}'. \apop(\Sigma(\vx)) \to \vv{d}' \para \cfc(E, \Sigma[\vy \mapsto \vv{d}'], t, \vv{d}, \vv{r})
\end{gather*}

Note that if $\apop$ is a function, the static restrictions we put on \lzero (\autoref{subsec:backend-assumptions}) force this case to be a non-recursive call, and it is subsequently handled in the linking stage by essentially substituting in the subgraph compiled from the body of $\apop$.

\paragraphb{Compiling Branching}
Dataflow operators execute whenever their inputs are available, so \system needs to ensure that the body of 
either branch does not directly use the channels from outside the branch. Otherwise, an operator in 
the ``wrong'' branch may be triggered. 
Therefore, \system inserts \apsteer operators as gates between live variables outside the branch and their uses inside each branch, so that the variables only flow to the path triggered by the branch condition. Naturally, after the \apsteer{}s and branch bodies, the compiler needs to combine the sets of outputs from both branches, which is done by three \apmerge operators, one for each output 
channel.
Formally, the definition is as follows, where $\vv\Sigma$ means treating $\Sigma$ as a list of channels indexed by live variables, and $\vv{\Sigma_1}, \vv{\Sigma_2}$ are two lists of fresh channel names for each variable in $\Sigma$:
\[
\begin{array}{l}
    \cfc(\mathbf{if}~x~\mathbf{then}~E_1~\mathbf{else}~E_2, \Sigma, t, \vv{d}, \vv{r}) :=
        \nu~\vv{\Sigma_1}, \vv{\Sigma_2}, t_1, t_2, \vv{d_1}, \vv{d_2}, \vv{r_1}, \vv{r_2}. \\
    \begin{array}{lr}
        \apsteer_\top(\Sigma(x), \vv\Sigma) \to \vv{\Sigma_1} \para
          \apsteer_\bot(\Sigma(x), \vv\Sigma) \to \vv{\Sigma_2} \para & \text{(Steer live variables)} \\
        \cfc(E_1, \Sigma_1, t_1, \vv{d_1}, \vv{r_1}) \para
          \cfc(E_2, \Sigma_2, t_2, \vv{d_2}, \vv{r_2}) \para & \text{(Compile branch bodies)} \\
        \apmerge_\mathsf{init}(\Sigma(x), t_1, t_2) \to t \para & \text{(Merge tail condition)} \\
        \apmerge_\mathsf{init}(\Sigma(x), \vv{d_1}, \vv{d_2}) \to \vv{d} \para \apmerge_\mathsf{init}(\Sigma(x), \vv{r_1}, \vv{r_2}) \to \vv{r} & \text{(Merge return and tail call)}
    \end{array}
\end{array}
\]

\paragraphb{Compiling a Return or Tail Recursion}
Returning in \lone means sending values to the output channels ($\vv{d}$), while a tail call means sending the tail call arguments back to the arguments of the current function through back edges ($\vv{r}$).
Additionally, these processes should set the tail condition $t$, and send dummy values to the unused channels ($\vv{r}$ in the case of return, and $\vv{d}$ in the case of a tail call) so that \apmerge operators compiled from any branching proceed correctly.
In other words: 
\begin{align*}
    \cfc(\mathbf{ret}(\vx), \Sigma, t, \vv{d}, \vv{r}) &:=
        \apconst_\bot \to t \para
        \apconst_{\mathbf{unit}} \to \vv{r} \para
        \apforward(\Sigma(\vx)) \to \vv{d} \para
        \apsink(\vv{\Sigma} \setminus \vx) \\
    \cfc(\mathbf{tail}(\vx), \Sigma, t, \vv{d}, \vv{r}) &:=
        \apconst_\top \to t \para
        \apconst_{\mathbf{unit}} \to \vv{d} \para
        \apforward(\Sigma(\vx)) \to \vv{r} \para
        \apsink(\vv{\Sigma} \setminus \vx)
\end{align*}

Unlike the case of a normal non-recursive function call ($\cfc(\mathbf{let}~\vy = \apop(\vx);~E)$), our static restrictions (\autoref{subsec:backend-assumptions}) do \emph{not} prevent multiple tail calls in the same function.
This is acceptable because if we have multiple tail calls in different branches of a function, the tail call arguments will eventually be merged by the $\apmerge$ gates compiled in the branching case above, and routed back to the start of the body with back edges.

This also alludes to the reason why we put the static restrictions in \lzero in the first place.
Within the same function body, the condition for merging arguments to a function is exactly the tail condition $t$, since the arguments can either come from an external caller, or from any of the tail calls within the function.
However, if we do allow multiple external call sites to a function, the conditions for merging these external sets of arguments are not \emph{local}, and we must sacrifice the compositionality of the compiler definition to perform more complex global analyses.

\paragraphb{Forward Simulation}
We now sketch our proof of forward simulation of the CFC pass (the full version has been formalized in Lean), which intuitively means that the output dataflow graph has at least one schedule that behaves the same as the input \lzeroghost function.

The main difficulty of establishing a simulation relation for this translation is that 
a single use of a variable in \lzeroghost can explode into a tree of channels in \lone. This 
is largely due to steering in a dataflow graph, where
any live variable needs to be routed to the correct branch, and the results of both branches need to be merged at the end. As an example, consider the simple conditional expression below: the variable $z$ has 
three uses, but the compiled dataflow graph has at least 12 channels (those named $c_{z_i}$) that receive the value of $z$ at some point in the execution!
\begin{align*}
    \centering
    & E \equiv \letin{z}{\apop(x, y)}{\ite{z}{\ret(x, z)}{\ret(y, z)}} \\
    &\cfc(E, \emptyset, t, \vv{d}, \vv{r}) = \apop(c_{x}, c_{y}) \to c_{z_1} \para
             \apfork(c_{z_1}) \to c_{z_2} \dots c_{z_{10}} \para & \text{(Run \apop and copy $z$)} \\
    & \ \ \apsteer_\top(c_{z_2}, c_{x}) \to c_{x'} \para
             \apsteer_\top(c_{z_3}, c_{z_4}) \to c_{z_{11}} \para & \text{(Route to true branch)} \\
    & \ \ \apsteer_\bot(c_{z_5}, c_{y}) \to c_{y'} \para
             \apsteer_\bot(c_{z_6}, c_{z_7}) \to c_{z_{12}} \para \dots \para & \text{(Route to false branch)} \\
    & \ \ \apmerge(c_{z_8}, \dots) \to \vv{d} \para \apmerge(c_{z_9}, \dots) \to \vv{r} \para \apmerge(c_{z_{10}}, \dots) \to t & \text{(Merge outputs)}
\end{align*}

To alleviate the issue and make the simulation proof tractable, we make two important design decisions.
First, we enforce affinity (i.e., no re-use) of variables and require no shadowing of variable names (\autoref{sec:typing}), essentially requiring \lzeroghost programs to be in an ``affine'' SSA form~\cite{dragon-book}.
This ``well-formedness'' restriction greatly reduces the complexity of tracking the liveness of a variable. 

Given this restriction, the second part of the solution is to use a consistent naming scheme in the compiler definition, so that each use of a variable $x$ in the source program corresponds to a \emph{unique} channel name that in essence tracks provenance.

Following is a snippet of our Lean definition of channel names.
\begin{lstlisting}[style=lean4, frame=none, aboveskip=3pt, belowskip=3pt, numbers=none]
inductive ChanName X where | var (base : X) (pathConds : List (Bool x ChanName X))
                           | steer_cond (chan : ChanName X) -- Other 10 cases omitted
\end{lstlisting}
The parameter \texttt{X} is the type of variable names in the source program, and after CFC, all variables in the compiled \lone process have type \texttt{ChanName X}.
The most common case of channel names is the \texttt{var} case, where \texttt{base} is the original variable name, and \texttt{pathConds} keeps track of the sequence of branching decisions required to reach the program point at which the variable occurs. Path conditions can in turn inductively contain \texttt{ChanName}, because the branching conditions themselves may be ``gated'' by other branching conditions.
Other cases such as \texttt{steer\textunderscore{}cond} are used when the variable is the decider of a \apsteer, \apmerge, etc.
In total, we use 12 cases of \texttt{ChanName} to disambiguate all channel names in the produced dataflow graph.

With the SSA form and channel naming scheme, the rest of the simulation proof is relatively standard.
The simulation relation keeps track of the current path condition (a list of branching conditions used and their Boolean value) and of the key correspondence between the local variable store of $\interp{\lzeroghost}$ and channel buffers of $\interp{\lone}$.
The simulation proof shows that this relation is preserved after one step of the input \lzeroghost function $\interp{D}$ and potentially multiple steps of $\interp{\cfc(D)}$.

\begin{theorem}
    \label{thm:cfc-sim}
    For a well-formed \lzeroghost definition $D$, $\interp{D} \iosim \interp{\cfc(D)}$. Furthermore, the initial state of $\interp{D}$ is only related to the initial state of $\interp{\cfc(D)}$ in the simulation.
\end{theorem}

The second part of the statement guarantees that after $D$ returns, $\cfc(D)$ restores to its initial state, which ensures deadlock-freedom even when the function is called multiple times.

\begin{figure}[tb]
    \small
\[
\begin{array}{c}
%
\inferrule*[right=Base]{
    c_0 \xrightarrow{l}_0 c'_0 \qquad
    l \in \{ \yieldl{\apop, \vv{v_{in}}, \vv{v_{out}}} \mid \apop \in O \} \cup \{ \inputl{\vv{v}}, \outputl{\vv{v}}, \tau \}
}{
    (0, c_0, c_1, \dots, c_n) \xrightarrow{l} (0, c'_0, c_1, \dots, c_n)
} \\

\\



\inferrule*[right=Spawn]{
    c_0 \xrightarrow{\yieldl{f_i, \vv{v_{in}}, \vv{v_{out}}}}_0 c'_0 \quad
    c_i \xrightarrow{\inputl{\vv{v_{in}}}}_i c'_i
}{
    (0, c_0, \dots, c_i, \dots) \xrightarrow{\tau} (i, c_0, \dots, c'_i, \dots)
} \qquad

\inferrule*[right=Ret]{
    c_0 \xrightarrow{\yieldl{f_i, \vv{v_{in}}, \vv{v_{out}}}}_0 c'_0 \quad
    c_i \xrightarrow{\outputl{\vv{v_{out}}}}_i c'_i
}{
    (i, c_0, \dots, c_i, \dots) \xrightarrow{\tau} (0, c'_0, \dots, c'_i, \dots)
}
\end{array}
\]
    \caption{Selected transition rules of $\linksem(S, S_1, \dots, S_n)$ for linking a base $S \in \lts{C}{\labelset_{O \sqcup \{ f_1, \dots, f_n \}}}$, and dependent LTSs $\{ S_i \in \lts{C_i}{\labelset_O} \}^n_{i = 1}$. Note that only the base LTS can ``call'' the dependent LTSs. The configuration of the linked LTS is $\mathbb{N} \times C \times C_1 \times \dots \times C_n$, with the first index denoting the currently active LTS.}
    \Description{Transitions of semantic linking.}
    \label{fig:link-sem}
\end{figure}

\subsection{Linking}
\label{sec:compiler-linking}
We can now lower a sequential \lzeroghost function to a parallel \lone process that is guaranteed to simulate the source semantics. However, the compiler still has not handled 
calls to other functions, and instead treats them as uninterpreted synchronous operators. 
This intentional design decision drastically improves proof modularity; 
if \system performs linking as a part of CFC, the final simulation relation has to relate the entire call stack of an \lzeroghost program to the channel buffers of \lone, which complicates the already involved channel naming scheme in \autoref{sec:compiler-cf}.
Instead, by splitting the two phases, the proofs in \autoref{sec:compiler-cf} only need to focus on a single call frame and the corresponding dataflow subgraph, leaving linking as a separate reasoning layer, which we explain now.

To compile the whole program, we must reason about function linking in both \lzeroghost and \lone, in a way that provably connects their semantics. Thus, \system uses two notions of linking for this pass:
(1) \emph{semantic} linking $\linksem$, which gives semantics to function calls in \lzeroghost;
and (2) \emph{syntactic} linking $\linksyn$, which ``stitches'' compiled \lone processes together into a single \lone process.

\paragraphb{Semantic Linking}
The high-level idea of semantic linking is to take the semantics (i.e., LTS) of each individual function definition (with uninterpreted function symbols), and construct a single LTS in which each component function can call one another. We leverage 
the fact that we enforce an acyclic call graph in \lzeroghost programs (\autoref{sec:overview}) to have a (relatively) simple definition where a \emph{base} LTS can make calls to a list of \emph{dependent} LTSs without mutual recursion.
We show a snippet of our definition in \autoref{fig:link-sem}, which is partially inspired by \emph{interaction semantics} in compositional CompCert~\cite{comp-compcert}, 
though more relaxed due to our call graph restrictions.

The definition revolves around a base LTS $S$ that can call operators in $O$ as well as function symbols $f_1, \dots, f_n$, 
which correspond to dependent LTSs $\{ S_i \}^n_{i = 1}$ that only call operators in $O$. The 
linked LTS, denoted $\linksem(S, S_1, \dots, S_n)$, is then an LTS that only calls operators in $O$, with function calls ``internalized'' as silent steps.
This linked LTS has its own transition relation. Intuitively, this relation treats all LTSs as parallel threads, with the current active thread indicated by the first index in the state. 
If a thread produces a label that does not involve interaction between the threads, the label is simply passed through. However, if the base LTS has a potential transition to produce $\yieldl{f_i, \vv{v_{in}}, \dots}$, and a dependent LTS $S_i$ can accept an $\inputl{\vv{v_{in}}}$ label, then they synchronize and the active thread switches to $S_i$ (\textsc{Spawn}). Once $S_i$ finishes execution and produces an $\outputl{\vv{v_{out}}}$ label, the base LTS resumes execution (\textsc{Ret}).
One point to note is that no component LTS ever resets to its initial state $\mathsf{init}_i$. While the state of a function ``resetting'' after returning makes sense in 
the sequential \lzeroghost setting, this is not generally true in the \lone semantics (even though it is satisfied by a process produced by CFC).
Therefore, we define the linked semantics this way to avoid dependency on the simulation proof of CFC and preserve modularity.

As an example of semantic linking, suppose we have the following two simple functions
\[
\mathbf{def}~f(x) = g(x) + 1; \mathbf{def}~g(x) = x + 1
\]
The LTS $\interp{f}$ can have labels $\yieldl{g, \vv{v_\mathrm{in}}, \vv{v_\mathrm{out}}}$ and $\yieldl{+, \vv{v_\mathrm{in}}, \vv{v_\mathrm{out}}}$, while the LTS $\interp{g}$ can only generate $\yieldl{+, \vv{v_\mathrm{in}}, \vv{v_\mathrm{out}}}$ (besides input/output/silent labels).
Then $\linksem(\interp{f}, \interp{g})$ would act like the \emph{parallel composition} of the two LTSs that interact when a call happens: whenever $\interp{f}$ yields to (calls) $g$, an input label will be sent to $\interp{g}$; whenever $\interp{g}$ returns, the return values are passed back to $\interp{f}$ through the yield label $\yieldl{g, \vv{v_\mathrm{in}}, \vv{v_\mathrm{out}}}$ and the execution of $f$ continues.

In this semantic version of linking, we prove a congruence lemma of $\linksem$ with respect to $\iosim$.
\begin{lemma}
    \label{lem:link-sem-congr}
    Given two sets of LTSs $S_0, \dots, S_n$ and $S'_0, \dots, S'_n$ such that $S_i \iosim S'_i$ for all $i = 0, \dots, n$, $\linksem(S_0, \dots, S_n) \iosim \linksem(S'_0, \dots, S'_n)$.
\end{lemma}

\paragraphb{Syntactic Linking}
We can use \autoref{lem:link-sem-congr} to prove forward simulation 
of semantically linked processes, i.e., $\linksem(\interp{D_1}, \dots, \interp{D_n}) \iosim \linksem(\interp{\cfc(D_1)}, \dots, \interp{\cfc(D_n)}).$
This result has a crucial gap, however, as the final compilation result of a function should be a single dataflow graph, not 
a collection of function graphs.
Therefore, we define a \emph{syntactic linking} operation for \lone processes, denoted $\linksyn(\pi, \pi_1, \dots, \pi_n)$, that replaces any operator call to $f_i$ in $\pi$ with the corresponding process $\pi_i$.
This is defined inductively on an \lone process, and we show the key rule below, where $\pi_i(\vv{c}) \to \vv{d}$ denotes replacing the free input (output) channels of $\pi_i$ with $\vv{c}$ ($\vv{d}$).
\[
\linksyn((\apop(\vv{c}) \to \vv{d}), \pi_1, \dots, \pi_n) := \begin{cases}
    \pi_i(\vv{c}) \to \vv{d} & \text{if } \apop = f_i \\
    \apop(\vv{c}) \to \vv{d} & \text{otherwise}
\end{cases}
\]

To connect the semantic and syntactic notions of linking, we prove the following lemma saying that a syntactically linked \lone process includes the behavior of semantically linking the processes.
\begin{lemma}[Linking Simulation]
    \label{lem:link-sem-syn}
    $\linksem(\interp{\pi}, \interp{\pi_1}, \dots, \interp{\pi_n}) \iosim \interp{\linksyn(\pi, \pi_1, \dots, \pi_n)}$
\end{lemma}

\paragraphb{Final Simulation}
By connecting \Cref{thm:cfc-sim} and \Cref{lem:link-sem-congr,lem:link-sem-syn}, we can now state our final forward simulation theorem from \lzeroghost programs to \lone processes, which shows that the combined passes of CFC and linking produce a dataflow graph that \emph{includes} the behavior of the \lzeroghost program.

For an \lzeroghost program $P = D_1, \dots, D_n$, we say that it is \emph{well-formed} if each definition $D_i$ is well-formed (i.e., affine variable usage and no shadowing of variable names), and $P$ satisfies the static restrictions in \autoref{subsec:backend-assumptions}.
We denote the semantics of $P$ as $\interp{P} := \linksem(\interp{D_1}, \dots, \interp{D_n})$ and the final compilation pass $\comp(P) := \linksyn(\cfc(D_1), \dots, \cfc(D_n))$.
\begin{theorem}[Compiler Forward Simulation]
    \label{thm:compiler-sim}
    For a well-formed program $P$, $\interp{P} \iosim \interp{\comp(P)}$.
\end{theorem}

\subsection{Guarded Semantics and Determinacy}
\label{sec:compiler-guarded}
The final forward simulation (\autoref{thm:compiler-sim}) shows that the produced dataflow graph will have \emph{at least one} schedule that behaves the same as the source \lzeroghost program.
However, due to the non-deterministic execution of 
\lone processes and shared memory between operators, execution
may diverge from this ``good'' schedule and produce different results or
termination behaviors. 

In other words, we need to prove \emph{determinacy}: assuming termination, the compiled
program produces the same result regardless of the schedule. 
To establish this, we re-introduce ghost permission tokens from
\autoref{subsec:ghost-lowering}. We add a
permission-checking layer to both \lzeroghost and \lone semantics that verifies that operators are called with
correct permissions, getting stuck on violations.
This approach enables us to prove two key properties: (1) if one ``good'' terminating trace of an \lone process is found satisfying the permission specification, then the \lone process is strongly normalizing in \emph{all} schedules;
(2) compilation preserves permission-satisfying, terminating traces, so that it is sufficient to check the permissions at the \lzeroghost level (\autoref{sec:typing}).

\paragraphb{Guarded Semantics}
Given a permission specification for all operators
(\autoref{subsec:ghost-lowering}), we want to have a semantic ``permission
checker'' that checks whether each operator call satisfies the permission
specification and gets stuck otherwise.
As a first step, we need a ``hook'' 
to reason about permissions within our semantics, so we slightly augment our operator set $O$ to create 
an extended set $\ext(O)$. The key difference of $\ext(O)$ is that 
each operator is given an additional input and output channel specifically for
ghost permission tokens (mirroring \lzeroghost's structure in \autoref{subsec:ghost-lowering}),
so that permission tokens will be included in the $\mathsf{yield}$
labels produced by both \lzeroghost and \lone semantics. Additionally, $\ext(O)$
contains the ghost operator $\join$ in \autoref{subsec:ghost-lowering}, 
which is used to split and join permission tokens.

Now that we have a semantics with permission tokens threaded through, we can create a checker to make 
sure that any operator call satisfies the permission spec. To do so, we define a 
notion of a \emph{guarded} semantics, which is in essence an LTS with a restricted and translated label set.

\begin{definition}[Guarded Semantics]
    \label{def:guarded-sem}
    Given a relation $G \subseteq \labelset_{O'} \times \labelset_O$ which we call \emph{guard}, and an $S = (\mathsf{init}, \to) \in \lts{C}{\labelset_{O'}}$, $\guard_G(S) = (\mathsf{init}_G, \to_G) \in \lts{C}{\labelset_O}$ is another LTS such that $\mathsf{init}_G := \mathsf{init}$, and for any $c, c' \in C$, $c \xrightarrow{l}_G c'$ iff there exists $l'$ such that $c \xrightarrow{l'} c'$ and $(l', l) \in G$.
\end{definition}

\begin{figure}[tb]
    \small
\[
\begin{array}{c}
\inferrule*[right=Guard-Yield]{
    t_{\mathit{pre}} = \precond(\apop, \vv{v_{in}}) \qquad
    t_{\mathit{post}} = \postcond(\apop, \vv{v_{in}}, \vv{v_{out}})
}{
    \yieldl{\apop, (\vv{v_{in}}, t_{\mathit{pre}}), (\vv{v_{out}}, t_{\mathit{post}})} \guardrel_{\spec} \yieldl{\apop, \vv{v_{in}}, \vv{v_{out}}}
} \\[12pt]

\inferrule*[right=Guard-Join]{
    f(\vv{v}) \cdot t' = t_1 \cdot t_2 \cdot \dots \cdot t_k
}{
    \yieldl{\join_{k, f}, (t_1, \dots, t_k, \vv{v}), (f(\vv{v}), t')} \guardrel_{\spec} \tau
} \qquad

\inferrule*[right=Other]{
    l \in \{ \inputl{\vv{v}}, \outputl{\vv{v}}, \tau \}
}{
    l \guardrel_{\spec} l
}
\end{array}
\]
    \caption{Semantic guard for interpreting $\ext(O)$ with respect to a specification $\spec = (\precond, \postcond)$. This relation interprets the extended operator set $\ext(O)$ 
    as operations in the base operator set $O$.}
    \Description{Semantic guard.}
    \label{fig:spec-guard}
\end{figure}

We now define a guard relation $\guard_\spec(S)$ in \autoref{fig:spec-guard} for checking the operator calls ($\yieldl{\dots}$) against the permission specification.
In the \textsc{Guard-Yield} rule, we check that the additional permission token input and output of the operator $\apop$ satisfy the permission specification, and translate the label back to the base operator set $O$.
In the \textsc{Guard-Join} rule, we take the sum of the input permissions and then split it as specified by the $\join_{k, f}$ operator, where $k$ is the number of input permissions, and $f$ determines how the two output permissions should be split, potentially depending on input values $\vv{v}$.
Since this is a ghost operator, it produces no external yields and emits a $\tau$ label.
Finally, in the \textsc{Other} rule, we pass through any other labels unchanged.

To reason about the base unguarded semantics, we also define a trivial guard $\guard_\triv$ (omitted for space), which is similar to $\guard_\spec$ but skips all permission checks.

\paragraphb{Strong Normalization}
With our permission checker, we can prove properties about how the guarded semantics of \lone processes relate to the unguarded semantics.
Namely, if we can find a trace in the guarded semantics that leads to a final state in both semantics, then even without 
the guards, any execution trace in the unguarded semantics will be ``eventually consistent'' with the same final state.
Furthermore, the unguarded semantics in this case will be strongly normalizing, as all traces will be bounded by the length of the terminating, guarded trace.

\begin{theorem}[Guarded Weak Normalization Implies Unguarded Strong Normalization]
    \label{thm:strong-norm}
    Let $c \in \configlone$ be an \lone configuration, and let $\to_\spec$ and $\to_\triv$ 
    be the guarded/unguarded semantics of \lone, respectively.
    If there exists a trace $c \xrightarrow{\tau^k}_\spec c_\bot$ (of $k$ steps) such that $c_\bot$ is final with respect to $\to_\triv$, 
    then for any trace $c \xrightarrow{\tau^{k_1}}_\triv c'$ in the unguarded semantics, there exists a trace $c' \xrightarrow{\tau^{k_2}}_\triv c_\bot$ in the unguarded semantics to the same final state $c_\bot$, and $k_1 + k_2 = k$.
\end{theorem}

Note that the statement only considers $\tau$ labels.
This is because in a step omitted from the paper for brevity, we concretely interpret all previously uninterpreted operator calls, so the resulting semantics only generates input/output/silent labels but no yield labels.

\paragraphb{Connecting with Forward Simulation}
We have now established good properties at two ends of the compilation pipeline: (1) via the type system in \autoref{sec:typing}, we know that well-typed \lzero programs can be elaborated to \lzeroghost programs that will satisfy the permission checker; (2) via \autoref{thm:strong-norm}, we know that if the compiled \lone process satisfies the permission checker and has a ``good'' terminating trace, then it is strongly normalizing.
To bridge the gap between (1) and (2), we use the following key congruence lemma that preserves
the \emph{modularity} between forward simulation and determinacy.

\begin{lemma}
    \label{lem:guard-congr}
    For any guard relation $G$ and LTSs $S$ and $S'$, $S \iosim S'$ implies $\guard_G(S) \iosim \guard_G(S')$.
\end{lemma}

Therefore, by carrying a terminating, guarded trace of \lzeroghost programs (guaranteed by the type system) through the forward simulation proofs and applying \autoref{thm:strong-norm}, we can prove the following determinacy result, which says that the compiled \lone process $\comp(P)$ is strongly normalizing and, in particular, produces the same output values $\vv{v_{out}}$ as the input \lzeroghost program $P$.

\begin{theorem}[Compiler Determinacy]
    Let $P$ be a well-formed \lzeroghost program and $\comp(P)$ its compiled \lone process. Suppose that there is a terminating trace in the guarded \lzeroghost semantics: 
    \[
        \mathsf{init}(P) \xrightarrow{\inputl{\vv{v_{in}}}}_\spec c_1 \xrightarrow{\tau^k}_\spec c_2 \xrightarrow{\outputl{\vv{v_{out}}}}_\spec c_\bot
    \]
    Then there exists $k' \in \mathbb{N}$, such that:
    for any $c'_2$ with $\mathsf{init}(\comp(P)) \xrightarrow{\inputl{\vv{v_{in}}}\ \tau^{k_1}}_\triv c'_2$, we have
    $ c'_2 \xrightarrow{\tau^{k_2}\ \outputl{\vv{v_{out}}}}_\triv c'_\bot$
    with $k_1 + k_2 + 2 = k'$ and $c_\bot$ and $c'_\bot$ have the same final memory state.
\end{theorem}

\subsection{Operator Selection and Optimization}
\label{sec:compiler-opt}

As the final step in \system, we implement a term rewriter on the produced \lone process to lower ``pseudo-operators'' to actual hardware operators and optimize for a smaller graph size.
In particular, any ghost permission tokens become simple control signals, and a \join is lowered to an \aporder operator and a \apfork operator, which together wait for all inputs and then output two control signals.

In total, we have 54 rewrite rules, and they are currently unverified.
The formal verification of these rewrites is an interesting problem on its own (\autoref{sec:related}).
For instance, we have the rewrite rule
$\apmerge(c_0, c_1, c_2) \to d \para \apsteer_\top(c_0, d) \to d' \Rightarrow \apforward(c_2) \to d'$,
i.e., if a \apmerge and a \apsteer have the same decider ($c_0$), then $c_2$ can be directly forwarded to the final output $d'$.
This rewrite may not preserve the original behavior if, for example, $c_2$ is ready, but the decider $c_0$ is always empty.
Hence, the correctness of this rule relies on the liveness property that
any value will eventually be consumed, which is implied by the termination assumption (\Cref{thm:strong-norm})
and our simulation relation that enforces empty channels in the final dataflow configuration.


\newcommand{\numeval}{10\xspace}

\newcommand{\evalcgra}{
\begin{tabular}{l|rrrr|rrrr|rrrr}
 & \multicolumn{4}{c|}{Steps} & \multicolumn{4}{c|}{Graph Sizes} & \multicolumn{4}{c}{\%CF Ops} \\
Test & \textbf{Wv} & \textbf{Wv}$^*$ & RP & RP$^*$ & \textbf{Wv} & \textbf{Wv}$^*$ & RP & RP$^*$ & \textbf{Wv} & \textbf{Wv}$^*$ & RP & RP$^*$ \\
\hline
\texttt{vadd} & 81 & 234 & 28 & 75 & 20 & 60 & 7 & 11 & 70 & 90 & 29 & 36 \\
\texttt{norm} & 182 & 559 & 53 & 174 & 23 & 70 & 7 & 11 & 74 & 91 & 29 & 36 \\
\texttt{relu} & 128 & 396 & 36 & 99 & 20 & 88 & 7 & 11 & 70 & 94 & 43 & 45 \\
\texttt{pool} & 903 & 1694 & 354 & 360 & 157 & 338 & 39 & 50 & 88 & 95 & 69 & 64 \\
\texttt{fc} & 876 & 2391 & 647 & 855 & 60 & 183 & 26 & 34 & 73 & 92 & 54 & 53 \\
\texttt{conv} & 793 & 1340 & 612 & 613 & 212 & 417 & 49 & 53 & 89 & 95 & 57 & 57 \\
\texttt{dmv} & 281 & 711 & 192 & 247 & 44 & 126 & 21 & 29 & 75 & 91 & 57 & 55 \\
\texttt{dmm} & 1912 & 2963 & 519 & 695 & 128 & 245 & 42 & 54 & 83 & 91 & 71 & 67 \\
\texttt{dither} & 538 & 1050 & 250 & 260 & 58 & 188 & 22 & 30 & 79 & 94 & 59 & 57 \\
\texttt{sort} & 8233 & 12615 & 986$^\times$ & 1364 & 166 & 399 & 45 & 52 & 84 & 95 & 64 & 62 \\
\end{tabular}
}

\newcommand{\evalhls}{
\begin{tabular}{l|rrr|rrr|rrr|rrr}
 & \multicolumn{3}{c|}{Cycles} & \multicolumn{3}{c|}{Exec.\ Time ($\mu$s)} & \multicolumn{3}{c|}{LUTs} & \multicolumn{3}{c}{FFs} \\
Test & \textbf{Wv} & \textbf{Wv}$^*$ & CRT & \textbf{Wv} & \textbf{Wv}$^*$ & CRT & \textbf{Wv} & \textbf{Wv}$^*$ & CRT & \textbf{Wv} & \textbf{Wv}$^*$ & CRT \\
\hline
\texttt{vadd} & 158 & 228 & 108 & 2.5 & 4.4 & 1.8 & 787 & 1372 & 979 & 667 & 1171 & 818 \\
\texttt{norm} & 385 & 555 & 258 & 7.0 & 10.8 & 4.3 & 1590 & 2284 & 1491 & 1007 & 2057 & 1086 \\
\texttt{relu} & 212 & 390 & 144 & 3.2 & 7.6 & 2.3 & 700 & 1763 & 1142 & 600 & 1492 & 954 \\
\texttt{pool} & 696 & 1231 & 662 & 25.2 & 51.7$^\dagger$ & 16.7 & 6444 & 10296 & 7960 & 6476 & 11040 & 7868 \\
\texttt{fc} & 1144 & 2065 & 1362 & 32.8 & 44.6 & 37.0 & 2368 & 4568 & 5285 & 1847 & 4034 & 4539 \\
\texttt{conv} & 678 & 955 & 764 & 23.4 & 46.4$^\dagger$ & 22.2 & 8851 & 13045 & 7599 & 8865 & 13748 & 6537 \\
\texttt{dmv} & 364 & 614 & 390 & 6.0 & 12.0 & 8.0 & 1797 & 3452 & 3902 & 1479 & 3002 & 3670 \\
\texttt{dmm} & 1522 & 2161 & 1326 & 54.0 & 75.4 & 31.3 & 5232 & 7263 & 6710 & 5074 & 6928 & 6513 \\
\texttt{dither} & 541 & 836 & 500 & 12.0 & 20.1 & 11.0 & 1845 & 4228 & 4634 & 1392 & 3415 & 4523 \\
\texttt{sort} & 5717 & 8833 & 5202 & 192.4 & 229.6 & 187.9 & 4840 & 8215 & 6004 & 4288 & 7531 & 5054 \\
\end{tabular}
}

\newcommand{\evalperf}{
\begin{tabular}{l|rrr|rrrr}
& \multicolumn{3}{c|}{Source LoC} & \multicolumn{4}{c}{Compiler Time (s)} \\
Test & C & DSL & Ann & TC & TV & Opt & Total \\
\hline
\texttt{vadd} & 5 & 26 & 2 & 0.03 & 0.67 & 0.02 & 0.72 \\
\texttt{norm} & 5 & 27 & 2 & 0.02 & 0.33 & 0.03 & 0.39 \\
\texttt{relu} & 7 & 30 & 3 & 0.02 & 0.39 & 0.06 & 0.47 \\
\texttt{pool} & 27 & 135 & 5 & 0.06 & 1.07 & 1.31 & 2.45 \\
\texttt{fc} & 16 & 71 & 4 & 0.09 & 1.76 & 0.23 & 2.08 \\
\texttt{conv} & 36 & 179 & 5 & 0.11 & 3.94 & 2.94 & 7.00 \\
\texttt{dmv} & 9 & 47 & 3 & 0.07 & 1.64 & 0.09 & 1.80 \\
\texttt{dmm} & 16 & 85 & 4 & 0.10 & 2.41 & 0.55 & 3.06 \\
\texttt{dither} & 19 & 75 & 7 & 0.09 & 1.26 & 0.30 & 1.66 \\
\texttt{sort} & 23 & 117 & 12 & 0.04 & 1.77 & 1.97 & 3.79 \\
\end{tabular}
}

\newcommand{\evalloc}{
\begin{tabular}{l|rrrr}
    Component & Time & Code & Spec & Proofs \\
    \hline
    Type Checker        & 8   &   8,113 &       - &       - \\
    Perm. Validator     & 4   &   3,433 &       - &       - \\
    CF Conversion       & 8   &     644 &     713 &   6,353 \\
    Linking             & 4   &     153 &      96 &   2,604 \\
    Determinacy         & 8   &       - &     725 &   5,862 \\
    Optimization        & 3   &     853 &       - &       - \\
    \hline
    Total               & 35  &  13,196 &   1,534 &  14,819 \\
\end{tabular}
}

\newcommand{\WvStepsOverRP}{2.62}
\newcommand{\WvNooptStepsOverRP}{5.79}
\newcommand{\WvStepsOverRPNoopt}{1.69}
\newcommand{\WvNooptStepsOverRPNoopt}{3.74}
\newcommand{\WvGSOverRP}{3.04}
\newcommand{\WvNooptGSOverRP}{8.26}
\newcommand{\WvGSOverRPNoopt}{2.26}
\newcommand{\WvNooptGSOverRPNoopt}{6.13}
\newcommand{\WvCycOverCRT}{1.12}
\newcommand{\WvNooptCycOverCRT}{1.78}
\newcommand{\WvExecOverCRT}{1.2}
\newcommand{\WvNooptExecOverCRT}{2.04}
\newcommand{\WvLUTsOverCRT}{0.69}
\newcommand{\WvNooptLUTsOverCRT}{1.22}
\newcommand{\WvFFsOverCRT}{0.67}
\newcommand{\WvNooptFFsOverCRT}{1.27}
\newcommand{\DSLOverC}{4.86}
\newcommand{\simProofRatio}{5.6}
\newcommand{\rustLoc}{11,546\xspace}
\newcommand{\leanCodeLoc}{797\xspace}
\newcommand{\leanSpecLoc}{1,534\xspace}
\newcommand{\leanProofLoc}{14,819\xspace}
\newcommand{\cfcProofLoc}{6,353\xspace}
\newcommand{\detProofLoc}{5,862\xspace}
\newcommand{\leanProofRatio}{6.4}
\newcommand{\WvCFPct}{78}
\newcommand{\WvNooptCFPct}{93}
\newcommand{\RPCFPct}{51}
\newcommand{\RPNooptCFPct}{52}

\section{Evaluation}
\label{sec:eval}

In this section, we evaluate \system in terms of compilation quality (\autoref{sec:eval-quality}), ergonomic/compiler overhead of \lzero (\autoref{sec:eval-overhead}), and human effort in verifying \system (\autoref{sec:eval-verif-effort}).
All evaluations are performed on a machine with Apple M1 Pro CPU and 32~GiB of RAM.

\begin{figure}[t]
  \centering
  \includegraphics[width=\textwidth]{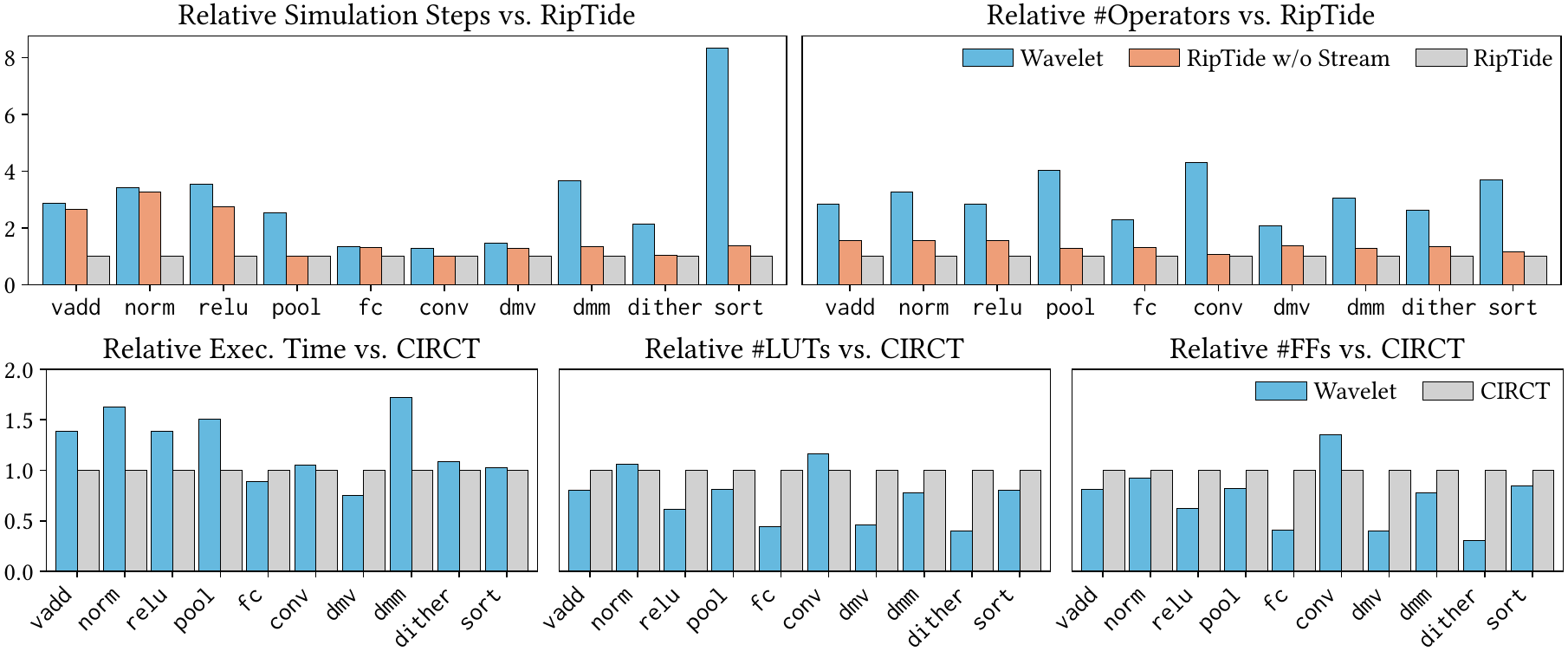}
  \caption{Compilation quality: the first row compares simulation performance and graph size of \system and RipTide; the second row compares execution time and resource usage of \system and CIRCT in dynamic HLS.}
  \Description{Compilation quality comparison.}
  \label{fig:eval-plots}
\end{figure}

\subsection{Compilation Quality}
\label{sec:eval-quality}

\system targets CGRAs such as RipTide~\cite{riptide} or FPGAs as a part of a dynamically-scheduled high-level synthesis (HLS) pipeline~\cite{circt,dynamatic}.
Therefore, we compare the performance and resource usage of dataflow graphs compiled from \system and two unverified dataflow compilers from RipTide~\cite{riptide} and LLVM CIRCT~\cite{circt} (for HLS) on 10 benchmark programs from RipTide.

\paragraphb{Comparison with RipTide}
The RipTide project~\cite{riptide} features an LLVM-based dataflow compiler targeting the RipTide CGRA.
The output dataflow graphs of \system and the RipTide compiler are at a similar level of abstraction, so we directly compare their simulation performance and graph sizes in the first row of \autoref{fig:eval-plots}.
Simulators of \system and RipTide are both not cycle-accurate (every operator takes one cycle), but their relative performance is still informative.

Compared to RipTide, \system's dataflow graphs are \WvStepsOverRP$\times$ slower and \WvGSOverRP$\times$ larger in geometric mean.
Without optimizations in \autoref{sec:compiler-opt}, \system's dataflow graphs are \WvNooptStepsOverRP$\times$ slower and \WvNooptGSOverRP$\times$ larger.
One advantage of RipTide is that it has a ``streamification'' optimization to replace a loop header similar to \autoref{fig:df-ops} (a) with a specialized ``stream'' operator, which we do not yet support in \system.
With streamification disabled in RipTide for a fairer comparison, \system's dataflow graphs are \WvStepsOverRPNoopt$\times$ (\WvNooptStepsOverRPNoopt$\times$ if unoptimized) slower and \WvGSOverRPNoopt$\times$ (\WvNooptGSOverRPNoopt$\times$ if unoptimized) larger.

The main downside of \system is that it produces significantly more control-flow operators (e.g., \apmerge, \apsteer, etc.).
In the benchmarks, \system-compiled graphs have \WvCFPct\% control-flow operators on average (\WvNooptCFPct\% if unoptimized), compared to \RPCFPct\% in RipTide's graphs.
We attribute this issue to the memory synchronization strategy in
\system.
\system passes permission tokens \emph{separately} from concrete values,
extending all operators with additional permission token inputs/outputs
(\autoref{sec:compiler-guarded}).
While this makes the type system more principled---permission variables are
affine and do not interfere with ordinary variables---it introduces many permission
variables and larger graphs.

On the other hand, RipTide \emph{mixes} synchronization signals with values and optimizes
away unnecessary memory ordering when data dependencies already enforce them.
However, this analysis is error-prone and harder to verify compositionally.
In fact, in the \texttt{sort} benchmark, we find that RipTide is missing a memory ordering, causing a data race and incorrect simulation results.

Finally, pipelining in \system does improve performance even with larger graphs.
For example, in benchmarks \texttt{vadd}, \texttt{norm}, and \texttt{fc}, even though \system produces graphs with $2\times$ as many operators as RipTide (without streamification), the final performance is still similar due to pipelining.

\paragraphb{Comparison with CIRCT}
CIRCT~\cite{circt} is an MLIR-based compiler framework for hardware design.
We compare \system against a dynamic HLS pipeline in CIRCT, which compiles MLIR's structured control-flow dialect (\texttt{scf}~\cite{mlir-scf}) to dataflow graphs (CIRCT's \texttt{handshake} dialect~\cite{circt-handshake}).
For \system, we implement an unverified lowering pass from \lone to the \texttt{handshake} dialect.

We then compile the produced \texttt{handshake} dataflow graphs to SystemVerilog RTL designs via CIRCT, and use Verilator~\cite{verilator} for simulation.
To estimate cycle periods and resource utilization, we synthesize the designs targeting the Lattice ECP5 FPGA~\cite{lattice-ecp5} using Yosys~\cite{yosys} and nextpnr~\cite{nextpnr}.

The resulting comparison is shown in the second row of \autoref{fig:eval-plots}, including relative execution time and resource utilization of the generated designs from \system and CIRCT.
The results show that \system's dataflow graphs, when used in HLS, are comparable with CIRCT's compilation results in both performance and area, with some benchmarks even outperforming CIRCT's output.
On average, \system's results are \WvExecOverCRT$\times$ (\WvNooptExecOverCRT$\times$ if unoptimized) slower in execution time, and have \WvLUTsOverCRT$\times$ (\WvNooptLUTsOverCRT$\times$ if unoptimized) LUTs and \WvFFsOverCRT$\times$ (\WvNooptFFsOverCRT$\times$ if unoptimized) FFs compared to CIRCT's output.

During evaluation, we actually found two issues in CIRCT~\cite{circt-pr-9587,circt-issue-9807}, which further highlight the challenge of building a correct dataflow compiler.

\begin{wrapfigure}{R}{0.57\columnwidth}
  \evalperf
  \caption{Benchmark lines of code (in the \textbf{C} version from RipTide~\cite{riptide}, in \system's \textbf{DSL}, and for capability \textbf{Ann}otations), along with their compilation time in various passes of \system{}: type checking (\textbf{TC}), translation validation (\textbf{TV}), optimization (\textbf{Opt}). Omitted CFC and linking took less than 10~ms.}
  \Description{Benchmark and compilation time statistics.}
  \label{fig:eval-compiler-perf}
\end{wrapfigure}

\subsection{\texorpdfstring{\lzero}{L*let} Type System Overhead}
\label{sec:eval-overhead}

To guarantee dataflow determinacy and to enable modular compiler proofs, our frontend language \lzero adopts a capability type system (\autoref{sec:typing}) and two static restrictions (\autoref{subsec:backend-assumptions}).
In this section, we evaluate both the compiler-performance and ergonomic overhead of the type system.

\autoref{fig:eval-compiler-perf} shows statistics of our benchmarks and compiler performance.
In terms of usability, benchmarks written in our \lzero DSL have on average \DSLOverC$\times$ more lines of code than the original versions in C.
While the ratio is high, this is primarily due to the lack of automated desugaring passes in our frontend.
Our DSL enforces individual operations in separate statements to aid compilation (e.g., \texttt{x\,=\,a\,+\,b\,*\,c} is written as \texttt{t\,=\,b\,*\,c;\,x\,=\,a\,+\,t}), and the two static restrictions in \autoref{subsec:backend-assumptions} require users to manually convert their control flow to branching and tail recursion.

Our capability types (\autoref{sec:typing}) place a relatively small burden on the programmer.
\autoref{fig:eval-compiler-perf} shows that users only have to add a few lines of annotations.
While precise capabilities and fences do require careful memory reasoning, this
overhead is reasonable given the strong assurance from \system.

Most compilation time is spent on translation validation (\autoref{subsec:backend-assumptions}) and dataflow optimizations (\autoref{sec:compiler-opt}).
For translation validation, some test programs with more complex array index computation can lead to inefficient non-linear integer arithmetic in the SMT queries.
For dataflow optimizations, the performance issue is largely due to inefficient internal representations of dataflow graphs.

\subsection{Verification Effort}
\label{sec:eval-verif-effort}

In total, the Lean formalization of \system has \leanCodeLoc lines of verified executable code, \leanSpecLoc lines of specifications, and \leanProofLoc lines of proofs, with a proof-to-code/spec ratio of about \leanProofRatio{}:1.
The type checker and permission validator are implemented in \rustLoc lines of Rust.

The most difficult proofs of \system are for control flow conversion ($\sim$2 person-months and \cfcProofLoc lines of proofs) and determinacy ($\sim$2 person-months and \detProofLoc lines of proofs).
Defining the channel naming schemes (\autoref{sec:compiler-cf}) and static restrictions (\autoref{subsec:backend-assumptions}) is the main difficulty of control flow conversion.
The determinacy proofs require careful coordination with the frontend in order
to achieve modularity, and the exact formulation of the guarded semantics
also took some iterations.

\section{Related and Future Work}
\label{sec:related}

The idea of dataflow is ubiquitous, and has been studied in various contexts
including dataflow architectures, stream processing~\cite{flo,mamouras-2020},
and synchronous data flow~\cite{sdf}.
We discuss related formal verification efforts in these areas
and highlight exciting future directions for \system.

\paragraphb{Dataflow and HLS}
FlowCert~\cite{flowcert} targets CGRA compiler correctness and inspired our modular determinacy proof.
The main difference is that FlowCert is entirely a translation validation approach
and does not provide the same level of formal guarantee.
\system also improves upon FlowCert with fine-grained capabilities, enabling more pipelining than FlowCert's simple permission tokens.

Our type system is inspired by Dahlia~\cite{dahlia},
but we extend memory capabilities with dependency on values, which is crucial for pipelining.
Vericert~\cite{vericert-hls} is a formally verified HLS compiler in Rocq~\cite{rocq},
although it cannot be directly applied to asynchronous dataflow due to static scheduling.

The Dynamatic~\cite{dynamatic} framework for dynamically-scheduled HLS
also uses asynchronous dataflow extensively.
Law et al.~\cite{dynamatic-model} formalize two Dynamatic-inspired dataflow models in Rocq.
Their dataflow calculus \textsf{Cilan} is an alternative to \lone, but they do not allow shared memory
between dataflow components, which simplifies their determinacy proof.
They also do not verify our main focus of dataflow compilation from a sequential program.
Graphiti~\cite{graphiti} and ElasticMiter~\cite{elastic-miter} are two prior efforts on
verifying dataflow graph rewrites for Dynamatic, which are complementary to our work.
Connecting the formalizations of \system and Graphiti would be an interesting future direction,
since our proof structure supports future extensions to some extent: rewrites with
verified forward simulation can be added after linking without affecting the determinacy proof (\autoref{lem:guard-congr}).

\paragraphb{Synchronous Data Flow}
The more restricted model of synchronous data flow~\cite{sdf} has been extensively studied, with verification
efforts including Kind 2~\cite{kind2}, a model checker for Lustre~\cite{lustre},
and V\'{e}lus~\cite{velus-1,velus-2}, a verified compiler in Rocq from Lustre to
CompCert's Clight~\cite{compcert}.
\emph{Asynchronous} dataflow in our case presents a
different set of challenges, e.g., determinacy and shared memory.

\paragraphb{Stream Processing}
Stream processing and functional reactive programming~\cite{functional-reactive-animation} are closely related paradigms.
Recent works such as Flo~\cite{flo} and stream types~\cite{stream-types} share similar goals,
including determinism and liveness,
but they target more abstract features like higher-order streams rather than
low-level concerns like shared memory, making
them not directly applicable to our challenges.

\paragraphb{Affine Types and Ghost Permissions}
Affine types and permission-based systems have been explored in languages such as
Rust~\cite{rust-lang} and Linear Haskell~\cite{linear-haskell}, and in various program logics and verification tools, such as 
Verus~\cite{verus,verus-sys}, Linear Dafny~\cite{linear-dafny}, and
Iris~\cite{iris}.
Our capability type system in \lzero is inspired by these prior works,
but tailored specifically for reasoning about memory access
ordering and pipelining in dataflow architectures.
Exploring richer capability types and more powerful program logics for dataflow would be an interesting future direction.

\paragraphb{Compiler Verification}
We build on ideas from a rich body of work on compositional compiler verification~\cite{comp-compcert,comp-cert-m,zhang-2024}.
Our definition of semantic linking is partially inspired by linking interactive semantics in compositional CompCert~\cite{comp-compcert} and interaction trees~\cite{itrees}.

\paragraphb{Handling Complex Control Flow}
The dataflow model formalized in this work is also known as ordered dataflow~\cite{riptide}, where channels have the semantics of FIFO queues.
Prior (unverified) work~\cite{riptide,dynamatic} in this model only handles single-function compilation.
In our case, the entire \lzero program with multiple functions can be thought of as a single program in SSA form~\cite{ssa-functional}.
Our approach of using the two static restrictions (\autoref{subsec:backend-assumptions}) is inspired by RipTide~\cite{riptide}, which normalizes the input SSA program to a form with nested loops and branching.
Another approach~\cite{circt,dynamatic} is to use non-deterministic merge operators, but they further complicate the determinacy proof due to the additional non-determinism~\cite{dynamatic-model}.
We leave for future work the verification of \emph{tagged dataflow}~\cite{ttda,tyr}, which can support more expressive control flow (e.g., higher-order functions in Id~\cite{ttda}).

\begin{acks}
We sincerely thank Joshua Gancher and Bryan Parno for their guidance on the early stage of this project.
We are grateful to Stephanie Balzer, Bas van den Heuvel, Peter Thiemann, Jan Hoffmann, and Cesare Tinelli for
discussions on deadlocks in asynchronous dataflow.
We thank Souradip Ghosh, Nathan Serafin, Mitchell Fream, Brandon Lucia, and Nathan Beckmann
for their help on the RipTide and Tyr CGRAs.
We thank the anonymous reviewers for their time and detailed feedback.
This work is supported in part by the \grantsponsor{GS100000001}{National Science Foundation}{http://dx.doi.org/10.13039/100000001} under Grant No.~\grantnum{GS100000001}{2403144}.
Any opinions, findings, and
conclusions or recommendations expressed in this material are those
of the authors and do not necessarily reflect the views of the
National Science Foundation.
\end{acks}

\section*{Data Availability Statement}

The source code and proofs of \system can be found in our GitHub repository:
\url{https://github.com/plum-umd/wavelet}.
Our Docker image for artifact evaluation~\cite{artifact} contains all scripts and instructions to
reproduce the evaluation results in \autoref{sec:eval}.

\bibliography{bib}

\end{document}